\documentclass[fleqn,usenatbib]{mnras}

\usepackage{newtxtext,newtxmath}

\usepackage[T1]{fontenc}

\DeclareRobustCommand{\VAN}[3]{#2}
\let\VANthebibliography\thebibliography
\def\thebibliography{\DeclareRobustCommand{\VAN}[3]{##3}\VANthebibliography}

\usepackage{graphicx}	
\usepackage{amsmath}	
\usepackage{hyperref}
\usepackage{xurl}

\usepackage{ulem}

\newcommand{\kms}{{\rm \,km\,s$^{-1}$}} 
\newcommand{\pc}{\,per\,cent}
\newcommand{\Sgrlon}{$\tilde{\Lambda}$} 
\newcommand{\Sgrlat}{$\tilde{B}$}
\defcitealias{Petersen21_LMCreflexmotionNature}{PP21}
\defcitealias{yaaqib24_Rashid}{Y24}
\defcitealias{Xue11_SDSSDR8BHBs}{X11}
\defcitealias{Belokurov2015colourMgrelationBHB}{BK15}

\title[The MW--LMC interaction with DESI BHBs]{Exploring the interaction between the MW and LMC with a large sample of blue horizontal branch stars from the DESI survey}

\author[A. Bystr\"om et al.]{Amanda Bystr\"om$^{1}$\thanks{E-mail: Amanda.Bystrom@ed.ac.uk},
Sergey E. Koposov$^{1,2}$,
Sophia Lilleengen$^{3}$,
Ting S. Li$^{4}$,
Eric Bell$^{5}$,
\newauthor
Leandro {Beraldo e Silva}$^{5,6}$,
Andreia Carrillo$^{7,3}$,
Vedant Chandra$^{8}$,
Oleg Y. Gnedin$^{5}$,
Jiwon Jesse Han$^{8}$,
\newauthor
Gustavo E. Medina$^{4}$,
Joan Najita$^{9}$,
Alexander H. Riley$^{3}$,
Guillaume Thomas$^{10}$,
Monica Valluri$^{5}$,
\newauthor
Jessica N. Aguilar$^{11}$,
Steven Ahlen$^{12}$,
Carlos Allende Prieto$^{13,10}$,
David Brooks$^{14}$,
Todd Claybaugh$^{11}$,
\newauthor
Shaun Cole$^{3}$,
Kyle Dawson$^{15}$,
Axel de la Macorra$^{16}$,
Andreu Font-Ribera$^{14,17}$,
\newauthor
Jaime E. Forero-Romero$^{18,19}$,
Enrique Gaztañaga$^{20,21,22}$,
Satya Gontcho A Gontcho$^{11}$,
\newauthor
Anthony Kremin$^{11}$,
Andrew Lambert$^{11}$,
Martin Landriau$^{11}$,
Laurent Le Guillou$^{23}$,
Michael E. Levi$^{11}$,
\newauthor
Aaron Meisner$^{9}$,
Ramon Miquel$^{24,17}$,
John Moustakas$^{25}$,
Francisco Prada$^{26}$,
Ignasi P\'erez-R\`afols$^{27}$,
\newauthor
Graziano Rossi$^{28}$,
Eusebio Sanchez$^{29}$,
David Schlegel$^{11}$,
Michael Schubnell$^{30,5}$,
David Sprayberry$^{9}$,
\newauthor
Gregory Tarl\'{e}$^{5}$,
Benjamin A. Weaver$^{9}$,
and Hu Zou$^{31}$\\
Affiliations are listed at the end of the paper.
}
\date{Accepted XXX. Received YYY; in original form ZZZ}

\pubyear{2025}

\begin{document}
\label{firstpage}
\pagerange{\pageref{firstpage}--\pageref{lastpage}}
\maketitle

\begin{abstract}

\noindent
The Large Magellanic Cloud (LMC) is a Milky Way (MW) satellite that is massive enough to gravitationally attract the MW disc and inner halo, causing significant motion of the inner MW with respect to the outer halo. 
In this work, we probe this interaction by constructing a sample of 9,866 blue horizontal branch (BHB) stars with radial velocities from the DESI spectroscopic survey out to 120 kpc from the Galactic centre. 
This is the largest spectroscopic set of BHB stars in the literature to date, and it contains four times more stars with Galactocentric distances beyond 50 kpc than previous BHB catalogues. 
Using the DESI BHB sample combined with SDSS BHBs, we measure the bulk radial velocity of stars in the outer halo and observe that the velocity in the Southern Galactic hemisphere is different by 3.7$\sigma$ from the North. 
Modelling the projected velocity field shows that its dipole component is directed at a point 22 degrees away from the LMC along its orbit, which we interpret as the travel direction of the inner MW. 
The velocity field includes a monopole term that is $-24$ \kms, which we refer to as compression velocity. 
This velocity is significantly larger than predicted by the current models of the MW and LMC interaction. 
This work uses DESI data from its first two years of observations, but we expect that with upcoming DESI data releases, the sample of BHB stars will increase and our ability to measure the MW–LMC interaction will improve significantly.

\end{abstract}




\section{Introduction}

The Large Magellanic Cloud\footnote{For a history of how the naming convention `Magellanic' came into practice, see \citet{Dennefeld20_LMChistory}. Some astronomers argue for a renaming of the Magellanic Clouds, see \url{https://physics.aps.org/articles/v16/152}.} (LMC) is the most luminous Milky Way (MW) satellite galaxy.
It has a distance to the Sun of about 50 kpc \citep{Pietrzynski19_LMCdistance}.
It was believed in the past that the LMC had completed several orbits around the MW \citep{Murai80_LMCmanypericentres}, but the current view is that it has just completed its first pericentre passage around the MW \citep{Besla07_LMCfirstpericentre, Kallivayalil13_LMCfirstpericentrepassage, Sheng24_LMCfirstpericentre}, though a second pericentric passage has not been completely ruled out \citep{Vasiliev24_LMCMWmodels}.

What mainly drove the change in understanding the past orbital history of the LMC was not only improved proper motion measurements \citep{Kallivayalil06_LMCfirstpericentre} but also an improvement of its mass estimate, since the uncertainty on the LMC mass leads to a range of orbital periods \citep{Kallivayalil13_LMCfirstpericentrepassage}.
Previous mass estimates have been as low as $(0.7-1.0) \times 10^{10}$ M$_\odot$ \citep{Hindman63_lowLMCmass} with an enclosed mass within 9 kpc up to $\sim 2 \times 10^{10}$ M$_\odot$ \citep{vanderMarel14_lowLMCmass}.
The current total LMC mass estimate is instead about an order of magnitude larger, in the range of $(1-2) \times 10^{11}$ M$_\odot$, based on several separate lines of evidence \citep[and references therein]{Vasiliev23_LMChaloreview}.
With such a large mass, the LMC displaces the MW disc MW halo \citep{Garavito-Camargo19_LMCinteractionmodels, Garavito-Camargo21_LMCBFE, Lilleengen23_OCstreamLMC}.
It perturbs orbits of halo objects due to its presence, e.g. stellar streams such as the Orphan-Chenab stream \citep{Koposov19,Erkal19_LMCOCperturbations, Koposov2023, Lilleengen23_OCstreamLMC}, the Sagittarius (Sgr) stream \citep{Vasiliev20_Tangoforthree}, and several other streams that had a recent close passage with the LMC \citep{Shipp19_LMCstreams, Shipp21_LMCstreams}.
It can also cause a warping of the MW's stellar and gaseous discs \citep{Laporte18_LMCMWdiskwarp}, and the total amplitude of the warp is a sum of both internal MW dynamics and the LMC's tidal influence \citep{Han23_tiltedMWdisk}.

Such a massive satellite galaxy is, more importantly, expected to directly affect its host, the MW.
Inspired by these newer LMC mass estimates, \citet{Gomez15_MWLMCinteractionmodel} modeled the interaction between the MW and the LMC in a first-infall scenario. They found that if the LMC has a total mass of at least $5 \times 10^{10}$ M$_\odot$, the approximation of an inertial Galactocentric reference frame is no longer valid. 
The MW will respond to the gravitational pull of the LMC as the position and velocity of the MW's centre of mass changes by up to 75 \kms\ in less than 0.5 Gyr.

As a satellite galaxy orbits the dark matter (DM) halo of its host, it is predicted to generate a density wake trailing behind it as it experiences dynamical friction \citep{Chandrasekhar43_DynamicalFriction}, which is induced even by smaller orbiting DM subhaloes \citep{Buschmann18_DMsubhalosdensitywake}.
\citet{Garavito-Camargo19_LMCinteractionmodels} constructed models of the LMC's interaction with the MW to predict the density and kinematic signatures induced by the DM wake of the LMC, and the ability of surveys to observe these signatures.
Projected on the observable sky, they identify an overdensity trailing the orbit of the LMC in the Galactic South, which is shifted further along a past point of the LMC's orbit the further out into the halo we go, as well as an overdensity in the Galactic North that persists at all distances.
They refer to these overdensities as the transient and the collective responses, respectively.
Both these overdensities have since been claimed to be observed by \citet{Conroy21_LMCoverdensities}, however \citet{Amarante24_haloanisotropy} are only able to identify the transient response.
The transient response has been linked to the Pisces overdensity, a previously known halo structure \citep{Watkins09_PiscesOverdensity} that is part of a stream dubbed the Pisces plume, which traces the past orbit of the LMC \citep{Belokurov19_Piscesplumelocalwake}.

\citet{Garavito-Camargo19_LMCinteractionmodels} identified expected signatures of the MW--LMC interaction in the kinematics of outer halo stars. 
Along the transient response they saw that the mean Galactocentric radial velocity is negative.
This is consistent with the linkage of the Pisces plume with the transient response where the Pisces plume also shows a net negative velocity \citep{Belokurov19_Piscesplumelocalwake}. 
\citet{Garavito-Camargo19_LMCinteractionmodels} also concluded that along the collective response the radial velocity is instead on average positive.
When Galactocentric distances increase, these radial motions increase in strength.
The authors interpreted this as a result of the MW's disc moving around the new orbital barycentre induced by the LMC, towards the LMC's pericentre, causing a so-called reflex motion in which the Galactic Northern sky is redshifted and the Southern is blueshifted.
This should appear like a dipole signal in radial velocity across the sky \citep{Petersen20_LMCreflexMWpotentialbias}.

This dipole signal occurs because the MW disc has shorter dynamical time-scales than the outer halo. 
Thus the disc will move as a whole in response to the LMC. 
The outer stellar halo is much slower to respond to the LMC and can be assumed to be fixed \citep{Petersen20_LMCreflexMWpotentialbias}.
Since the Solar system is situated in the MW disc and comoving with it in response to the LMC, to us this will make it seem like the halo is blueshifted in the Southern Galactic hemisphere, and redshifted in the North. 
The position on the sky where the stars are most blueshifted is the direction of the MW disc's movement.
For this reason we also expect a stronger amplitude in the projected velocities as we look further into the outer halo \citep[hereafter \citetalias{yaaqib24_Rashid}]{yaaqib24_Rashid}, as dynamical time-scales increase the further into the halo we go.
This movement of the MW disc means that the MW is not in dynamical equilibrium, and not accounting for this when estimating its mass can lead to mass biases \citep{Erkal20_LMCreflexMWmassbias, Brooks24_MWmassbias}.

The redshift of the Northern Galactic hemisphere and the blueshift of the Southern hemisphere was measured using K-giants, BHBs, blue stragglers (BSs) and RR Lyrae (RRLe) with Galactocentric distances larger than 50 kpc as halo tracers by \citet{Erkal2021LMCouterhalobulkvelocity}.
The velocities of the two hemispheres are found to differ by 3.8 $\sigma$ and the North is receding with $\sim 11$ \kms, and the South is approaching with $\sim -27$ \kms, meaning that the two hemispheres clearly have distinct velocities.
In an accompanying simulation, they could see that the larger the LMC mass, the higher are the velocity amplitudes of the two hemispheres.
The work by \citet[hereafter \citetalias{Petersen21_LMCreflexmotionNature}]{Petersen21_LMCreflexmotionNature} found, using K-giants, BHBs and MW satellites as halo tracers but with Galactocentric distances larger than 40 kpc, that the disc is travelling with a velocity of 32 \kms, directed towards the so-called apex direction, which is at $(l_\text{apex}, b_\text{apex}) = (53^\circ, -28^\circ)$.
This means that the disc is moving towards a past point in the LMC's orbit, due to the delay in the disc response, in the Galactic South.
The amplitude of the travel velocity, like the bulk velocities, also probe the LMC mass; the heavier the LMC, the larger the travel velocity.
Not only does the stellar halo show signs of the interaction between the MW and the LMC, but the MW's satellite galaxies do too.
\citet{Makarov23_satellitetracersLMChaloitneraction} observed that because of the LMC, MW satellite galaxies with Galactocentric distances less than 100 kpc experience a bulk motion with respect to the Sun.

\citetalias{yaaqib24_Rashid} expanded on the work by \citetalias{Petersen21_LMCreflexmotionNature}, using their compilation of K-giants and BHBs as halo tracers, by examining the behaviour of velocity and apex direction as a function of distance.
They saw that the amplitude of the velocity signal, i.e. the travel velocity, increases with distance, and that the direction of this travel changes with distance, such that it moves from pointing towards the Galactic North at Galactocentric distances between 20--30 kpc, to being in the Galactic South for distances between 30 kpc and beyond 50 kpc.
Both \citetalias{Petersen21_LMCreflexmotionNature} and \citetalias{yaaqib24_Rashid} used a velocity field model containing a free parameter for the average Galactocentric radial velocity of their tracer stars, which can be interpreted as the global inwards (if negative) or outwards (if positive) motion of the entire halo.
This average radial velocity was found to be negative, meaning that the halo's velocity field is globally blueshifted on top of the dipole signal, implying a compression of the halo.
This was suggested to be caused by the bound mass of the LMC that has fallen into the MW, that deepens the MW potential causing a net inward motion of the halo.

\citet{Chandra24_LMCMW} used the same model of the velocity field induced by the LMC as the previous two authors, but on data covering the entire sky. 
They use only RGB stars as their distant halo tracers, reaching as far as 160 kpc.
They were able to measure the change in the travel velocity of the disc and apex direction from 40 kpc out to 120 kpc Galactocentric distance.
In that distance range, the disc travel velocity increases from $\sim 5$ \kms\ to $50$ \kms. 
They also observed in the Southern hemisphere quadrant containing their apex direction that the mean velocity becomes more negative with distance, and that the radially opposing quadrant in the Northern hemisphere instead becomes more positive with distance:
the amplitude of the velocity field induced by the LMC increases the further out into the halo we go.

Identifying a sample of distant halo tracers is necessary to investigate the effect of the LMC on the outer MW halo. 
Blue horizontal branch (BHB) stars are excellent distant halo tracers, due to their high intrinsic luminosity and simple polynomial colour-absolute magnitude relation \citep{Deason11_BHBcolourmagrelation, Belokurov2015colourMgrelationBHB, Barbosa22_newBHBrelationsSEGUE}. 
Since one of the main signals of the MW being affected by the LMC is in radial velocity, it is crucial that these stars have accurate radial velocities. 

In this paper, we use the Dark Energy Spectroscopic Instrument (DESI) spectroscopic measurements of BHB stars with radial velocities to characterise the effect of the LMC on the MW.
DESI contains a targeted BHB program providing excellent tracers of the MW--LMC interaction because of DESI's depth of observation, allowing us to reach far into the halo, and gain highly accurate radial velocities for all of the target stars.
The DESI BHB sample is the largest spectroscopic BHB sample in the literature. 
It will let us reach 120 kpc into the halo using stars that all have accurate distances and radial velocities.
Unlike other works, we restrict ourselves to only using BHB stars as our halo tracers. 
This ensures that our distances are affected by the same selection effects across the entire distance range used in the analysis, as other tracers such as K-giants have less straightforward distances to derive. 
Because the absolute magnitude distributions are different for each tracer type, avoiding mixing tracer populations leave us with a sample of stars that have homogeneous distances across the entire distance range too.
RRLe are good distance indicators, but difficult to model due to their pulsations, making the velocities more complex to derive accurately.
BHBs on the other hand are straightforward to model.

In Sec. \ref{sec:data}, we describe the DESI BHB observations with radial velocities and how we construct a BHB sample that is cleaned from halo substructure.
In Sec. \ref{sec:halovelocityfield} we characterise the velocity field of the outer halo and compare our results to models in Sec. \ref{sec:modelcomparison}.
We discuss and contextualise our BHB sample and results in Sec. \ref{sec:discussion} and finally give a summary and present our conclusions in Sec. \ref{sec:summaryconclusions}.

\section{Data}\label{sec:data}

The analysis is carried out using DESI BHB stars. In this section, we introduce the DESI survey and its dedicated Milky Way survey and BHB targeting, and how we clean the BHB sample from contaminants and halo substructure and compute the stellar distances.

\subsection{The DESI spectroscopic survey}\label{sec:data_DESI}

The Dark Energy Spectroscopic Instrument (DESI) is a multi-object spectrograph designed for ground-based wide-field surveys that operates on the Mayall 4-meter telescope at Kitt Peak National Observatory.
Its goal is to be the largest spectroscopic survey to date with a footprint of 14,000 square degrees covered over five years of observations, starting in May 2021.
It has 5,000 fibers and covers a wavelength range of 360 nm to 980 nm with resolutions between 2,000 and 5,500 depending on the wavelength \citep{DESI22_DESIoverview}. 
DESI can obtain spectra of almost 5,000 objects simultaneously over a $\sim 3 ^\circ$ field \citep{DESI16_instrumentdesign, Silber23_DESIrobots, Miller23_DESIopticalcorrector}.
The survey will obtain spectra for approximately 40 million galaxies and quasars and 12 million stars \citep{DESIcollaboration16_DESIdesign}.

DESI's main mission is a galaxy and quasar redshift survey to probe baryonic acoustic oscillations and constrain the properties of dark energy by delivering the most precise measurement of the expansion history of the universe ever \citep{Levi13_Snowmass2013}.
These observations are made when the sky conditions are dark and transparency is good. 
However, during 440 hours per year, the sky brightness is higher due to the moon and twilight, and seeing is worse, making observations past $z=0.6$ difficult.
In these conditions, DESI switches to making observations for the Bright Time Survey, of which the Milky Way Survey (MWS) is one, along with the Bright Galaxy Survey \citep{DESIcollaboration16_DESIdesign}. 

DESI began its 5 month long survey validation campaign in December 2020 \citep{DESI24_DESISV}, and the scientific part of the survey began in May 2021.
In June 2023, DESI published its early data release (EDR)\footnote{\url{https://data.desi.lbl.gov/doc/releases/edr/}}, which consists of the survey validation data \citep{DESI23_EDR}.
The MWS published a stellar value-added catalogue based on this EDR that contains radial velocity and stellar parameter measurements for about 400,000 unique stars \citep{Koposov24_MWSEDR}.
The first data release (DR1)\footnote{\url{https://data.desi.lbl.gov/doc/releases/dr1/}} was published in March 2025 and contains observations taken between May 2021 through June 2022 \citep{DESI25_DESIDR1}, with a value-added catalogue published by the MWS containing over 6 million sources with radial velocities and stellar parameters \citep{Koposov25_MWSDR1}.
The BHB sample produced from DR1 data in this work was also made public as a value-added catalogue\footnote{\url{https://data.desi.lbl.gov/doc/releases/dr1/vac/mws-bhb/}}.
Since then, the survey has progressed and has as of writing completed 93 \pc\ of observations in dark conditions and 96 \pc\ of observations in bright conditions.

The work in this paper is based on DR1 \citep{DESI25_DESIDR1}, as well as spectra observed in the second year of DESI observations between June 2022 and June 2023 that are not yet prepared for public release.
This latter data will be referred to as year 2 data, or Y2. 
The DR1 dataset consists of coadditions (or coadds), i.e. averages, of spectra processed similarly to the EDR \citep{DESI23_EDR}.
In the EDR, spectra of objects in a given survey and program are coadded within HEALPixels on the sky using an $N_\text{side} = 64$ nested HEALPix tesselation. 
One HEALPixel can cover several tiles so that spectra of objects that were observed in different tiles are still coadded \citep{Guy23_DESIdataprocessing}.
The Y2 dataset is different, because it has been based on daily processing of DESI data and therefore consists of stacked spectra within individual DESI tiles \citep{Schlafly23_Dailyoperations}. 
This means that stars that were observed on multiple tiles will not have their spectra coadded.
In short, DR1 data is coadded across HEALPixels and Y2 across tiles.

Using DR1 \citep{DESI25_DESIDR1}, the DESI collaboration has released so-called key papers presenting the two point clustering measurements and validation \citep{DESI25_2ptclustering}, baryonic acoustic oscillations measurements from galaxies and quasars \citep{DESI25_BAO} and from the Lyman $\alpha$ forest \citep{DESI25_Lya}, as well as a full-shape analysis of galaxies and quasars \citep{DESI24_fullshapeanalysis}. 
There are cosmological results from the baryonic acoustic oscillations measurements \citep{DESI25_cosmoresultsBAO} and the full-shape analysis \citep{DESI25_cosmoresultsfullshape}.

\subsection{The DESI Milky Way Survey}

The MWS' science goals are to understand the chemical and dynamical evolution and star formation history of the MW’s thick disc and stellar halo, and identify the remnants of dwarf galaxies that have been accreted by the MW. 
This will be achieved by delivering radial velocities and chemical abundances for distant stars at high Galactic latitudes, targeting stars that are too faint for reliable velocities from other surveys such as Gaia.
At least 6.6 million unique spectra are expected to be observed during the survey's nominal five years.

The MWS has three main survey target categories: a) the  main sample, b) the high-priority sample consisting of white dwarfs, RRLe, BHBs and nearby stars, and c) the faint sample.
These are all observed within the bright-time program.
The main sample is split into MAIN-BLUE, MAIN-RED and MAIN-BROAD, whose union is a magnitude-limited selection within $16<r<19$, where $r$ is an extinction-corrected magnitude. 
From now on, all magnitudes mentioned are extinction-corrected.
MAIN-BLUE is defined by $(g-r) < 0.7$, and all stars redder than this belong to the MAIN-RED category if they satisfy additional astrometric requirements; if they are redder than this but do not pass those requirements, the stars belong to the MAIN-BROAD category.
The faint samples are all stars with $19<r<20$, and are also split into a blue and a red category.
See \citet{Cooper2023MWSoverview} for more details. 

The MWS provides radial velocities to all its targets via the radial velocity and stellar parameter fitting code RVSpecfit, which we will refer to as the RVS pipeline. 
This also includes all stellar secondary targets, spectrophotometric standards, or those targets that the spectroscopic pipeline Redrock\footnote{\url{https://github.com/desihub/redrock/}} (Bailey et al. 2024, in prep.) identified as a star. 
The underlying ideas for the RVS pipeline were presented in \citet{Koposov11_RVSapplication}: non-flux calibrated stellar spectra are fit by models, constructed by multiplying interpolated stellar templates by a polynomial to fit the continuum. 
This has been implemented in the Python package \texttt{RVSpecFit} \citep{Koposov19_RVS}, which is publicly available\footnote{\url{https://github.com/segasai/rvspecfit}}. 
It relies on the PHOENIX grid of synthetic spectra from \citet{huseer2013}.  
The code has been adapted to fit DESI spectra by computing the total log-likelihood function for each of the spectrograph's three arms \citep{Cooper2023MWSoverview}.  
The outputs from the pipeline are not only the best-fitting radial velocity, but also best-fitting stellar atmospheric parameters [Fe/H], log($g$) and [$\alpha$/Fe], and the stellar rotation $V \sin{(i)}$, and the uncertainties on these parameters.
These outputs are used to create a 64 bit warning bitmask called $\texttt{RVS\_WARN}$. 
$\texttt{RVS\_WARN}$ consists of four individual bitmasks, out of which we only consider the following three: \texttt{\{'CHISQ\_WARN': 1, 'RV\_WARN': 2, 'RVERR\_WARN': 4\}}.
These flags are set if a given spectrum's difference in $\chi^2$ of the stellar model, i.e. the stellar template multiplied by the polynomial, with respect to just the polynomial is smaller than 50, which indicates a non-stellar target; if the fitted radial velocity is within 5 \kms\ of the velocity interval edges at [--1500, 1500] \kms; or if the fitted radial velocity error is larger than 100 \kms.

The outputs and the quality of the RVS measurements based on the DESI EDR data are described in \citet{Koposov24_MWSEDR}. 
However, this paper relies on an updated version of RVS, that now uses a neural network as an interpolator between stellar templates. 
This new interpolation technique fixed the concentration of measured stellar parameters at grid points in metallicity, surface gravity and effective temperature seen in \citet{Koposov24_MWSEDR}. 

\subsection{BHB stars in the MWS}\label{sec:data_DESIBHBtargeting}

The MWS contains a high-priority sample.
It has higher priority than the main and faint samples, which in order of highest priority to lowest is a sample of white dwarfs, RRLe, nearby stars, and BHB stars.
These targets are prioritised when DESI assigns fibers for observations, to ensure high completeness.
There are in total 10,695 BHB targets included in DR1 and that thus are coadded by HEALPixel, and 10,498 in Y2, coadded by tiles.
Both within DR1 and within Y2, and in between DR1 and Y2, there are repeat observations of the same star, so these duplicates will be removed.
DR1 and Y2 BHBs also include both main survey and secondary targets, and we will introduce them separately below.

The selection of the main survey BHB stars in DESI is described in \citet{Cooper2023MWSoverview} and is based on the basic definition of main sample stars, but with additional criteria. 
Here we repeat these additional criteria for completeness:

\begin{itemize}
    \item $G>10$
    \item $\varpi \leq 0.1 + 3 \sigma_\varpi$ mas 
    \item $-0.35 \leq (g-r) \leq -0.02$
    \item $-0.05 \leq X_\text{BHB} \leq 0.05$
    \item $r - 2.3(g-r) - W1_\text{faint} < -1.5$
\end{itemize}

\noindent where $G$ is the Gaia magnitude, $\varpi$ is the Gaia parallax and $\sigma_\varpi$ the associated error, $g$ and $r$ are DESI Legacy Imaging Survey optical photometry, and

\begin{equation}\label{eq:mainsurveyBHBcolourselection}
\begin{aligned}
    X_\text{BHB} &= (r-z) - [1.07163(g-r)^5 - 1.42272(g-r)^4 \\
    & + 0.69476(g-r)^3 -0.12911(g-r)^2 \\
    & + 0.66993(g-r) - 0.11368],
\end{aligned}
\end{equation}

\noindent which is a boundary of BHBs and BSs in colour-colour space that is used to remove BS contamination from the BHB sample\footnote{Note that there is a typo in Eq. (1) in \citet{Cooper2023MWSoverview} where the first term of the equation is $(g-z)$, which is supposed to be $(r-z)$ as in this work.} (this relation is originally from the $S^5$ survey described in \citet{Li19_S5} and adopted for the MWS, see their Eq. (6) and the corresponding section for more details), and

\begin{equation}
    W1_\text{faint} = 22.5 - 2.5\log_{10}{(W1 - 3\sigma_{W1})}
\end{equation}

\noindent where $W1$ is the WISE 3.4 $\mu$m flux and $\sigma_{W1}$ its error.
There are in total 9,917 main survey BHB targets in DR1 and 9,840 from Y2. 

BHBs are not only observed as main survey targets by the MWS, but also as secondary targets.
Secondary targets are observed in the dark-time program, meaning that instead of the $16<r<20$ magnitude limit for the main survey BHB targets, secondary targets reach $19<r<21$.
Some targets belong to both target categories, but only those that are in the latter category are observed in dark time.
These targets are more distant than main survey targets observed in bright-time, helping us reach distances up to 120 kpc, which is not possible in bright-time observations. 
Secondary target BHBs are defined by the following criteria: 

\begin{itemize}
    \item $-0.3<(g-r)<0.0$
    \item $-0.7<(r-z)<-0.05$
    \item $-0.3 \cdot (g-r)^2 < ((r-z) - 0.75 \cdot ((g-r) + 0.15) + 0.27 - 0.025)< 0.06 - 0.6 \cdot ((g-r) + 0.15)^2)$
    \item $W1_\text{ratio} < 0.3 \cdot (g-r) + 0.15 + 3W1_\text{ratio,error} + 0.3$
    \item no saturation in the $g$-, $r$- or $z$-bands (or that none of \texttt{MASKBITS\footnote{\url{https://www.legacysurvey.org/dr9/bitmasks/}} ALLMASK\_G, ALLMASK\_R, ALLMASK\_Z} is set)
    \item none of the observed stars is on top of a galaxy from the atlas by \citet{Moustakas23_SGA} (\texttt{MASKBIT GALAXY} is not set) 
\end{itemize}

\noindent where 

\begin{equation}
    W1_\text{ratio} = \frac{W1}{g}
\end{equation}

\noindent is the extinction corrected flux ratio in W1 vs DeCALS $g$-band, and the error on this flux ratio is $W1_\text{ratio,error}$:

\begin{equation}
    W1_\text{ratio,error} = \frac{1}{\sqrt{W1_{\text{ivar}}} \cdot g}
\end{equation}

\noindent where $W1_{\text{ivar}}$ is the inverse variance of the W1 flux.

There are in total 1,453 secondary BHB targets in DR1 and 1,368 from Y2.
There is an overlap between the main survey and secondary targets of 675 stars in DR1 and 710 stars in Y2, which is why the total number of stars in DR1 and Y2 is 10,695 and 10,498 respectively, are smaller than the sum of their main survey and secondary targets.
There is also an overlap between DR1 and Y2 of 1,291 targets.
The total number of BHB targets, both main survey and secondary targets and across the data processing categories, is then 19,902 targets after duplicates have been removed.

\subsection{Cleaning the BHB sample}\label{sec:data_BHBselection}

The BHB target selection above is based only on photometry and astrometry, which will introduce contamination from several sources but mainly in the form of BS stars.
Here we present the steps taken to remove that contamination using DESI spectroscopy to create a sample of spectroscopically confirmed BHBs.

We begin by identifying objects that are not stars.
All objects that are flagged as quasars by the template-fitting code Redrock (Bailey et al. 2024, in prep.) are removed using the selection \texttt{RR\_SPECTYPE = QSO}.
What we are left with is a sample consisting of only stars that all have been run through the RVS pipeline to obtain their radial velocities. 
We make sure that the sample only contains good-quality radial velocities by removing stars that had bad radial velocity flags \texttt{CHISQ\_WARN}, \texttt{RV\_WARN} or \texttt{RVERR\_WARN} set.

In the resulting stellar sample, potential stellar contaminants are RRLe and BSs.
RRLe are good tracers of the halo and can be used for this analysis alongside the BHBs, but we choose to not include them as their pulsations make their spectra more difficult to model than BHBs and their velocities more complex to derive.
We remove the RRLe by crossmatching with the PanSTARRS1 3$\pi$ survey RRLe catalogue \citep{Sesar17_RRLe} and the Gaia third data release RRLe sample \citep{Clementini23_GaiaRRLe}.
This removes 167 RRLe.

BSs are stars that appear more massive and younger than stars of the same populations and thus appear bluer than stars of the same luminosity, because they formed through mergers or mass-transfer between two or three stars \citep{Perets09_BSorigin}.
BS have similar colours to BHBs, and so they contaminate simple BHB colour-based selections, as e.g. the MWS target selection for BHBs.
Eq. \eqref{eq:mainsurveyBHBcolourselection} for the main survey BHB targets and the requirements on both $(g-r)$, $(r-z)$ for the secondary BHB targets are designed to remove most of this BS contamination.
However to not compromise BHB completeness, these colour cuts are not very aggressive, so we wish to remove BS stars without removing more BHBs. 
These remaining BS can be removed by utilising spectroscopic information derived from the RVS pipeline, see e.g. \citet{Clewley02_BHBvsBS, Xue08_SDSSBHBSample}. 

\begin{figure*}
 \includegraphics{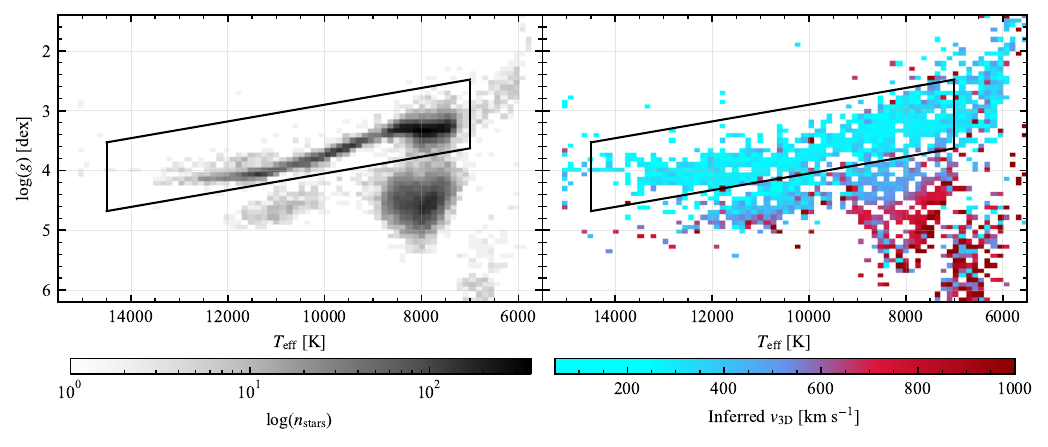}
 \caption{ 
 All DESI candidate BHB targets in Kiel space after removing quasars, stars with bad radial velocities, and RRLe, colour-coded by density (\textit{left}) and inferred three-dimensional velocity, (\textit{right}). 
 This figure includes the BHBs that are kept in the final sample, which are all stars within the black lines given by Eq. \eqref{eq:Kielcuts}. 
 The contamination outside of this box mainly comes from BS stars.
 The choice of the colour-coding of the right-hand panel follows the approach in the left-hand panel of Fig. 11 in \citet{Li19_S5} to understand the BS contamination.
 Assuming that all stars in the sample are BHBs, one can infer their distances using Eq. \eqref{eq:Mgcolourrelation} and calculate their three-dimensional velocities using these distances and their (correct) radial velocities.
 If these inferred three-dimensional velocities are too large, the assumption that the stars are BHBs is wrong as their distances have been overestimated.
 This allows us to identify BS stars, which are not as intrinsically luminous as BHBs and have higher surface gravities. They are visible as the red coloured area below the BHB sequence in the right-hand panel.
 }
 \label{fig:Kielspace}
 \end{figure*}

BSs are removed by making cuts in $T_\text{eff}$ and log($g$), otherwise known as Kiel space.
Fig. \ref{fig:Kielspace} shows the Kiel diagram of the sample after the quality cuts given previously in this section have been applied to it, colour-coded by density in the left-hand panel and by inferred three-dimensional velocity (assuming they are BHB stars) in the right-hand panel, together with the cuts we apply in this space as black lines.
These cuts are designed to keep only the BHB sequence extending across the figure, where we wish to mainly avoid the BSs that can be seen as the cloud of stars beneath the sequence at $\sim 8,000$ K.
We keep all stars with:

\begin{equation}\label{eq:Kielcuts}
\begin{aligned}
    7,000 < T_\text{eff} < 14,500 \text{ K} 
    \text{ and } 
    1.50 < \text{log(}g) - k \cdot T_\text{eff} < 2.65
\end{aligned}
\end{equation} 

\noindent where $k = 0.00014$.
We choose the range in $T_\text{eff}$ to avoid the cooler region of the horizontal branch where RRLe reside \citep{Giusti24_hotHB} and the edge of the temperature grid used in RVS, which is 15,000 K, even though the horizontal branch has been found to extend into temperatures as hot as 30,000 K \citep{Giusti24_hotHB}.

As we see from Fig. \ref{fig:Kielspace}, this keeps all stars that follow the BHB sequence in Kiel space, but mainly removes stars that have too high log($g$) or too low $T_\text{eff}$ values.
We also see from the right-hand side panel, that the stars we remove that do not lie on the BHB sequence, have relatively high inferred three-dimensional velocities, assuming BHB distances.
This confirms that they are BS contaminants, as contaminants will have overestimated distances if they are assumed to be BHB stars, but the measured radial velocities are correct; using both to compute the three-dimensional velocities will also yield overestimated velocities.

Once the BSs are removed, BHBs outside the colour range $-0.3 < g - r < 0.0$ are removed as that is the colour range in which we can derive reliable distances, see Sec. \ref{sec:data_distances}.
After the removal of RRLe stars and these quality cuts and cuts on surface gravity, effective temperature and colour are applied, the 19,902 BHB candidate targets have been reduced to a sample of 9,866 spectroscopically confirmed BHB stars.

\subsection{BHB distances}
\label{sec:data_distances}

An argument for using BHBs as tracers of the outer halo is that their absolute magnitudes, and thus distances, are easy to derive as they can be computed using only colours.
We use the colour-absolute magnitude relationship in the DES $g$-band, requiring the DES $(g-r)$ colour, from \citet[hereafter \citetalias{Belokurov2015colourMgrelationBHB}]{Belokurov2015colourMgrelationBHB}, which is only valid in the colour regime $-0.30 < (g-r) < -0.05$.
This means that restricting our sample to this colour range will remove redder, but luminous, BHBs that could be used to probe further into the outer halo.
For that reason, the polynomial is extended into the redder regime $-0.05 < (g-r) < 0.00$ with a linear equation.

The absolute magnitude-colour relationship in \citetalias{Belokurov2015colourMgrelationBHB} was derived using DES photometry calibrated to APASS survey \citep{Koposov2015}. 
Therefore we need to ensure that the distances computed using the distance modulus are accurate when applied to DESI Legacy Imaging Survey photometry.

The DESI BHB sample contains a large number of Sgr stream stars, which has been mapped with RRLe by \citet{hernitschek2017Sgrdistancesplines}.
They provide a spline for heliocentric distances and Sgr stream longitude \Sgrlon\ for the leading arm, that we interpolate to create a function of heliocentric distances as a function of \Sgrlon\ (depicted in orange in Fig. \ref{fig:Sgrcuts}).
In Sec. \ref{sec:data_Sgrflagging} we introduce the selection of Sgr stream members using this interpolated spline, and we apply this selection here as well to identify Sgr leading arm stars.
Then for all stars that we flag as belonging to the Sgr leading arm in the range $50^\circ <$ \Sgrlon\ $< 100^\circ$, we compute the distances derived using the relationship in \citetalias{Belokurov2015colourMgrelationBHB} $d_\mathrm{BK15}$, and divide them by the RRLe-inferred Sgr distances $d_\mathrm{Sgr}$. 
The fraction $d_\mathrm{BK15}/d_\mathrm{Sgr}$ is 1.080.
This means that the BHB distances derived by using the relationship in \citetalias{Belokurov2015colourMgrelationBHB} are overestimated by around 8 \pc, where we assume that the RRLe-derived distances are correct.
A part of this uncertainty might also come from the RRLe distances, but nevertheless, the colour-magnitude relation must be corrected.
The 8 \pc\ overestimation implies a distance modulus shift of $-5 \log_{10}(1.080) = -0.167$; to undo this shift, this value is subtracted from the \citetalias{Belokurov2015colourMgrelationBHB} formula and its linear extrapolation to compute $M_{g}$.
The final BHB absolute magnitude-colour relation we get is,

\begin{equation}\label{eq:Mgcolourrelation}
\begin{split}
    \text{in the} & \text{ range} -0.30 < (g-r) < -0.05: \\
    \\
M_g =& \text{ } 0.566 - 0.392(g - r) + 2.729(g - r)^2 \text{ } + \\  
    & \text{ } 29.1128(g - r)^3 +  113.569(g - r)^4, \\
    \\
    \text{and in} & \text{ the range} -0.05 < (g-r) < 0.00: \\
    \\
M_g =& \text{ } 0.564 - 0.500(g-r).
\end{split}
\end{equation}

We validate these distances against GCs next. 
We do this by taking stars from \citet{Vasiliev21_GCcatalog} with more than 90 \pc\ probability of being GC members, which were selected based on Gaia proper motions and parallaxes, and crossmatch with DECaLS DR9 to get their dereddened photometry.
BHB stars are photometrically selected as GC members with $-0.3 < (g-r) < 0$ and $ -0.15 < (r-z) - (g-r) < 0.02$, which yields 803 BHBs in 23 GCs. 
Using the GC literature distances from \citet{Vasiliev21_GCcatalog} for each star we compute its absolute magnitude $M_{g, \mathrm{GC}}$, and computing the absolute magnitude $M_{g}$ using the DECaLS magnitudes in Eq. \eqref{eq:Mgcolourrelation}, we can compute the distance modulus offset $\Delta M_g = M_{g} - M_{g, \mathrm{GC}}$.
This distribution shows that the photometric selection contains outliers that likely are BSs, so we make an additional cut in $|\Delta M_g| < 1$, which yields 722 BHBs in 22 GCs. 
We find that the median of $\Delta M_g$ is $0.093$ mag, which implies a systematic distance underestimation of about 4.2 \pc, and the standard deviation is 0.133 mag.
This systematic offset is likely caused by a metallicity dependence in absolute magnitudes of BHBs \citep{Fermani13_BHBmetallicityanddistance}, which will be explored further in future work.

\begin{figure}
 \includegraphics{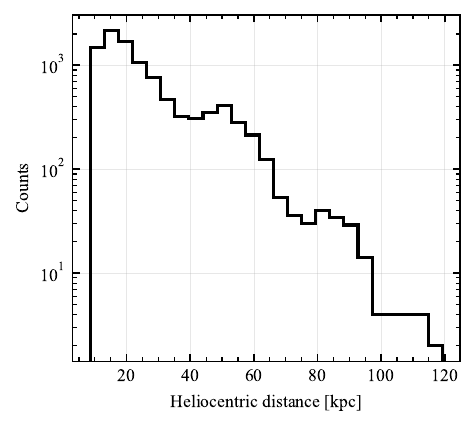}
 \caption{ 
 The heliocentric distance distribution of the DESI BHB sample after the BHB selections in Sec. \ref{sec:data_BHBselection} are applied to the parent sample. 
 The bumps around 50 and 85 kpc are Sgr stream stars, the removal of which is discussed in Sec. \ref{sec:data_Sgrflagging}.
 }
 \label{fig:heliocentricdistancedistribution}
 \end{figure}

The distance distribution of the DESI BHBs after applying Eq. \eqref{eq:Mgcolourrelation} is shown in Fig. \ref{fig:heliocentricdistancedistribution}. 
The maximum distance is 123 kpc and the distribution peaks around 15 kpc.
We also see a second peak at around 50 kpc and a third around 85 kpc, which are due to the Sgr stream.
The removal of these stars will be discussed in the next subsection, Sec. \ref{sec:data_Sgrflagging}.

\subsection{Removing halo substructure}\label{sec:data_Sgrflagging}

To characterise the velocity field of the halo, we need to remove known substructures in the halo such as GC or dwarf galaxy member stars or debris from past merger events that are not yet completely phase-mixed, such as stellar streams.
If left in the sample, these substructures could bias the velocity signal of the halo that we are looking for, as they each have their own distinct distribution of velocities.
GC stars are removed by crossmatching the sample with the catalogue by \citet{Baumgardt21_GCcatalog, Vasiliev21_GCcatalog}. 
Dwarf galaxy stars are removed by crossmatching the sample with high probability members from \citet{Pace22_dwarfgalaxystars}.
Both crossmatches use a 1 arcsecond radius.

The DESI observational footprint overlaps with the on-sky track of the Sgr stellar stream \citep{Cooper2023MWSoverview}, the biggest stellar stream in the halo. 
This is also reflected in a bump at about 55 kpc and one at about 85 kpc in Fig. \ref{fig:heliocentricdistancedistribution}: within the DESI footprint Sgr reaches from 20 to 120 kpc, covering the distances where we see these bumps.
We remove Sgr from our data in a similar way to \cite{Erkal2021LMCouterhalobulkvelocity}. 
First, we compute the Sgr longitude coordinates \Sgrlon\ and latitude coordinates \Sgrlat\ from \cite{Belokurov2014Sgrcoords} for all stars in the sample. 
Second, we compute which stars are within 15 degrees along \Sgrlat.
Third, we use the mean of the heliocentric distance distribution of Sgr as well as its line-of-sight depth as a function of \Sgrlon\ provided by \cite{hernitschek2017Sgrdistancesplines} to compute which stars are within 3 times the line-of-sight depth of the mean distance track of both the leading and trailing arm of Sgr. 
Combining these two conditions, all stars that pass the following criteria:

\begin{equation}\label{eq:Sgr_cut}
    |\tilde{B}| < 15^\circ
    \text{ and } 
    \left|\frac{\text{distance} - \text{distance}(\tilde{\Lambda})}{\sigma(\tilde{\Lambda})}\right|  < 3
\end{equation}

\noindent are labelled Sgr stars and removed from the analysed sample. 
This method identifies 1,854 Sgr stars. 
After the cuts to remove stars in GCs, dwarf galaxies, Sgr and Cetus-Palca \citep{Thomas22_Cetus-Palca} (see Sec. \ref{sec:data_Cetus-Palca}), the final sample used for the analysis is reduced from 9,866 to 7,960 BHBs. 

The combination of the selections guarantees a three-dimensional stream selection, as we combine the on-sky track given by (\Sgrlon, \Sgrlat) with the line-of-sight depth. 
Other works select Sgr using angular momentum (\citealp{Johnson20_ChandraLMCcut, Penarrubia21_Sgrangularmomentum}; \citetalias{Petersen21_LMCreflexmotionNature, yaaqib24_Rashid}); \Sgrlat\ only \citep{Thomas18_Sgrsel, Starkenburg19_Sgrspur}; \Sgrlon, distance and \Sgrlat\ as in this work, supplemented with cuts in velocity \citep{Erkal2021LMCouterhalobulkvelocity}; or a combination of cuts in all four aforementioned parameters \citep{Chandra24_LMCMW}.
We however do not choose to make additional selections in velocities or angular momentum along the stream, as to not bias the sample in velocity space since that is the space in which we will conduct our analysis. 
It also allows us to use velocity space as an additional check that the Sgr removal works, as we expect stars of a stream to have similar velocities at similar stream longitudes. 
 
\begin{figure*}
 \includegraphics{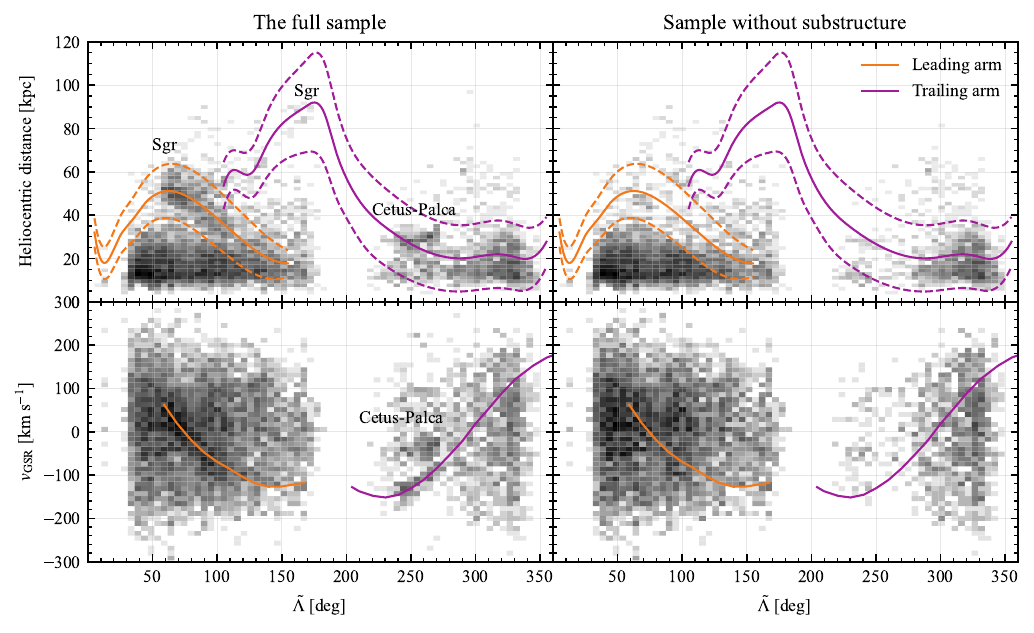}
 \caption{ 
 The sample before Sgr removal (\textit{left}) and after (\textit{right}), shown for heliocentric distance as a function of Sgr longitude coordinates \Sgrlon\ (\textit{top}) and for $v_\text{GSR}$ as a function of \Sgrlon\ (\textit{bottom}) with a Sgr velocity model by \citet{Vasiliev20_Tangoforthree}. 
 The Sgr leading (trailing) arm interpolated splines from \citet{hernitschek2017Sgrdistancesplines} are shown in solid orange (purple), with the corresponding interpolated 3$\sigma$ track shown in dashed orange (purple).
 The same colour convention is used for the velocity model by \citet{Vasiliev20_Tangoforthree}. 
 There are clear overdensities in the left-hand panel close to the leading arm apocentre around (\Sgrlon, distance) = (60$^\circ$, 60 kpc) and a smaller overdensity around (\Sgrlon, distance) = (250$^\circ$, 35 kpc), both of which is completely removed in the right-hand panel. 
 The latter overdensity is the Cetus-Palca stream \citep{Thomas22_Cetus-Palca}, see Sec. \ref{sec:data_Cetus-Palca}, which is also seen in the bottom left panel at about (\Sgrlon, $v_\mathrm{GSR}$) = (250$^\circ$, --50 \kms).
 To see the stars selected as Sgr members in the same space (i.e. those removed from the right-hand column), see Fig. \ref{fig:appendixSgrcolumnrownormalised}.
 } 
 \label{fig:Sgrcuts}
 \end{figure*}

The BHB sample before and after the Sgr removal can be seen to the left and right in Fig. \ref{fig:Sgrcuts} respectively, with the right-hand side of the figure clearly showing the Sgr overdensities disappearing after the selection has been applied.
The top row shows the stars' heliocentric distances as a function of \Sgrlon; it is in this space, combined with the third axis of \Sgrlat, that the Sgr selection is made.
The bottom row shows the radial velocity in Galactic standard of rest (GSR), $v_\text{GSR}$\footnote{\url{https://docs.astropy.org/en/latest/coordinates/example_gallery_rv_to_gsr.html} was used for computing $v_\text{GSR}$.},
as a function of \Sgrlon, which is an independent check that the removal of Sgr stars worked: Sgr in our sample clearly aligns with the Sgr velocity model from \citet{Vasiliev20_Tangoforthree}, and even though we make no explicit selections in velocity space, the stars on Sgr-like velocities are also removed from the sample.
In the upper left panel at \Sgrlon\ $\sim 60-100$ degrees, we see that the Sgr leading arm appears vertically extended, which can be explained by the bifurcation of the stream leading arm  \citep{Belokurov06_FieldofStreams}. 
The top right of the same figure shows the Sgr leading arm apocentre ((\Sgrlon, distance) $\sim(70^\circ$, 60 kpc)).
This apocentre has been observed in RRLe \citep{Sesar17_Sgrspur, hernitschek2017Sgrdistancesplines} and simulations \citep{Gibbons14_spurmodel} to contain a spur-like structure that is also present in our data and removed with our Sgr removal defined above.

\subsubsection{The Cetus-Palca stream}\label{sec:data_Cetus-Palca}

At around (\Sgrlon, distance) = (250$^\circ$, 35 kpc) in the upper left panel of Fig. \ref{fig:Sgrcuts}, a narrow overdensity perpendicular to the Sgr trailing arm is visible.
In the bottom left panel at the same \Sgrlon, there are two separate overdensities: one following the stream track, and one above it at less negative velocities.
As can be seen in the right panel of the figure, both of these overdensities are identified and removed using Eq. \eqref{eq:Sgr_cut}.
However, it seems like these are two distinct overdensities: one very clearly is Sgr, following the Sgr stream velocity tracks, while the other one has a different and distinct velocity distribution.
The perpendicularity of the stream to Sgr \citep{Koposov12_SgrsouthandCetus}, its $v_\text{GSR}$ being centered around -50 \kms\ \citep{Yuan19_Cetus}, and its location in (\Sgrlon, distance) being similar to the stellar distribution in (R.A., distance) in Fig. 6 in \citet{Thomas22_Cetus-Palca} with the same corresponding distance, is consistent with observations of the Cetus-Palca stream.
Due to its proximity in configuration and velocity space to Sgr, Eq. \eqref{eq:Sgr_cut} identifies it, as can be seen in the right-hand column of the figure.
We decide that this removal is enough to remove Cetus-Palca and we do not add additional cuts designed for this specific stream. 
If there are Cetus-Palca stars left in the sample after this cut, they will not affect the outer halo velocity field characterisation, as they have too small Galactocentric distances (Cetus-Palca can be found at 35 kpc with a small distance gradient that is not enough to reach Galactocentric distances beyond 50 kpc \citep{Thomas22_Cetus-Palca}, which is the distance regime that our analysis will focus on).
This will keep the sample as unbiased as possible.

\subsection{Final sample}\label{sec:data_finalsample}

The 19,902 DESI BHB candidate targets were reduced to 9,866 confirmed BHBs after spectroscopic cleaning. 
This is the final DESI BHB sample.
This sample was then reduced to 7,960 BHBs after substructure was removed.
A few of these stars lie 3$\sigma$ away from the sample's mean velocity so that they have velocities exceeding 311.20 \kms, and we remove these velocity outliers.
This reduces the number of BHB stars to 7,947.
Removing them does not affect the results of the paper, however it is beyond the scope of this paper to further characterise these stars.
This reduced DESI BHB sample, without substructure and velocity outliers, will be used for the scientific analysis in this work.

The reduced DESI BHB sample is then combined with the BHB sample from \citet[hereafter \citetalias{Xue11_SDSSDR8BHBs}]{Xue11_SDSSDR8BHBs}.
The distances given by Eq. \eqref{eq:Mgcolourrelation} are on average 4.5 \pc\ smaller than the SDSS BHB distances published in \citetalias{Xue11_SDSSDR8BHBs}, who use a different relationship than us to derive their BHB absolute magnitudes.
The difference however is well within their mean distance errors of 10 \pc\ and accounting for it does not affect the final results of this paper, so we keep the original SDSS distances for the \citetalias{Xue11_SDSSDR8BHBs} part of the sample. 
The same GC, dwarf galaxy, and Sgr identification criteria as introduced in Sec. \ref{sec:data_Sgrflagging} for the DESI sample are applied to \citetalias{Xue11_SDSSDR8BHBs} to remove halo substructure.
Additionally, all stars with SDSS colours $(u - g) < 1.15$ and $(g - r) > - 0.07$, as well as stars satisfying $(u - g) < 1.15$ and c($\gamma$) $< 0.925$, are removed from the sample to remove BS contamination following \citet{Lancaster19_SDSScolourcut}.
The DESI and \citetalias{Xue11_SDSSDR8BHBs} samples have 1,281 stars in common for which the DESI radial velocities are used as they have smaller measurement errors.
The combined final DESI and \citetalias{Xue11_SDSSDR8BHBs} BHB sample consists of 13,076 stars. 
After removing Sgr, Cetus-Palca, GC and dwarf galaxy stars, and also velocity outliers, from the \citetalias{Xue11_SDSSDR8BHBs} sample, the reduced combined, or the final, BHB sample contains 10,650 BHBs used for the analysis in this paper. 
It has 675 BHB stars with Galactocentric distances larger than 50 kpc, out of which 617 are DESI BHBs, making this an excellent sample to study the effects of the LMC on the outer halo.

\begin{figure}
 \includegraphics{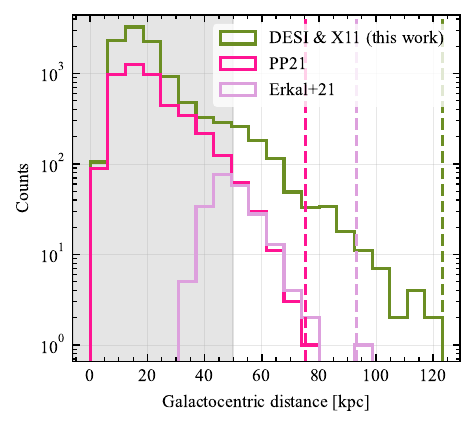}
 \caption{ 
 The Galactocentric distance distribution of our sample (green), the BHB sample used by \citetalias{Petersen21_LMCreflexmotionNature} (magenta) and the BHB sample used by \citet{Erkal2021LMCouterhalobulkvelocity} (lilac). 
 The sample in \citetalias{Petersen21_LMCreflexmotionNature} is based on the BHB sample by \citetalias{Xue11_SDSSDR8BHBs}, and the one used by \citet{Erkal2021LMCouterhalobulkvelocity} is a compilation of \citet{Xue08_SDSSBHBSample}, \citet{Deason12_BHBs} and \citet{Belokurov19_Piscesplumelocalwake}. 
 The corresponding dashed vertical lines mark the maximum distance of each sample. The grey shaded area to the left of 50 kpc marks the inner halo, and we only use the 675 stars past 50 kpc (non-shaded area) for our analyses.
 } 
 \label{fig:distancedistributions}
 \end{figure}

\begin{figure}
 \includegraphics[width=\columnwidth]{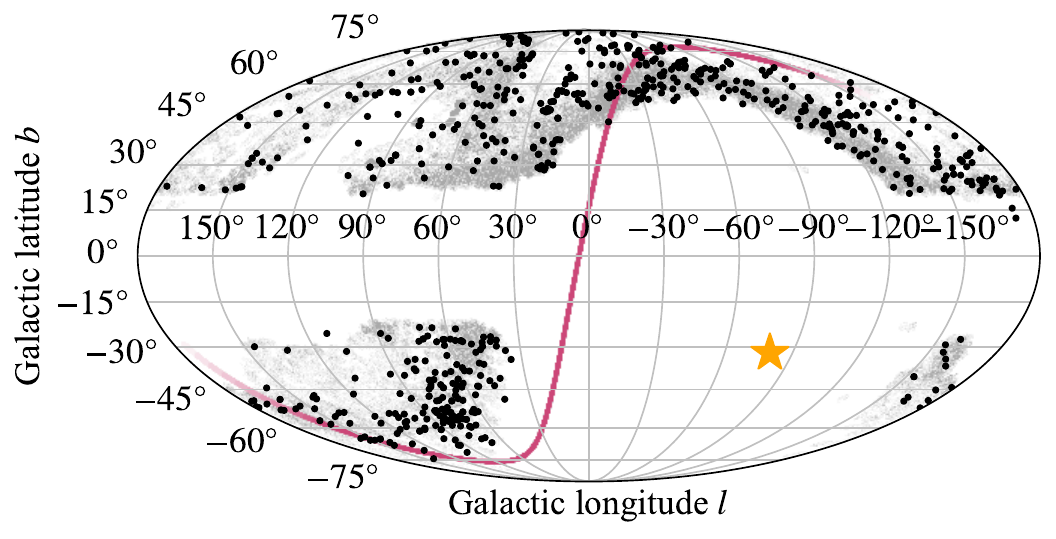}
 \caption{ 
 Distribution in Galactic coordinates of all the 675 stars in the final BHB sample that have Galactocentric distances larger than 50 kpc and thus are used for the scientific analysis, shown as black points.
 The current position of the LMC is shown as an orange star and all MWS targets observed as part of both the primary and secondary programs in DR1 and Y2 is shown in greyscale, to show the underlying footprint from which the BHBs come.
 Because DESI conducts its observations in circular tiles, the pattern on the sky looks like a large collection of dots; these dots do not represent one individual target, but an individual tile.
 The Sgr track from \citet{Mateu23_galstreams} is shown in purple.
 } 
 \label{fig:onskydistribution}
 \end{figure}

This is the largest spectroscopic BHB sample in literature which also contains the largest Galactocentric distances.
Fig. \ref{fig:distancedistributions} shows the distribution of Galactocentric distances of our final sample, compared to two other BHB samples from \citetalias{Petersen21_LMCreflexmotionNature} and \citet{Erkal2021LMCouterhalobulkvelocity}. 
These are based on the BHB samples by \citetalias{Xue11_SDSSDR8BHBs} (SDSS DR8) and \citet{Xue08_SDSSBHBSample} (SDSS DR6), \citet{Deason12_BHBs} and \citet{Belokurov19_Piscesplumelocalwake}, respectively. 
Our final sample's on-sky distribution is seen in Fig. \ref{fig:onskydistribution}, which shows that we cover larger portion of the Northern hemisphere than the Southern hemisphere, with a hole around the LMC.

\section{Characterising the halo velocity field}\label{sec:halovelocityfield}

To investigate the large-scale velocity patterns predicted in simulations \citep{Garavito-Camargo19_LMCinteractionmodels}, we will now study the radial velocity distribution of the BHBs in our sample as a function of Galactic coordinates and Galactocentric distance.

\subsection{Maximum likelihood estimation}

To characterise the velocity distribution of the halo, we need to analyse the average motions of the BHBs selected by their spatial positions. 
We will model the stars by assuming that their velocities can be described by a Gaussian distribution with a mean $\mu$ and a standard deviation $\sigma$. 
When evaluating the likelihood for each star, we also take individual uncertainties on the radial velocity measurements $\epsilon_i$, for each observed velocity $v_{r,i}$, into account.
We need to find the underlying velocity distribution that maximizes the likelihood of observing the measured velocities. 
Thus, we characterise the halo's velocity field using maximum likelihood estimation (MLE). 
The log-likelihood that we want to maximize is:

\begin{equation}\label{eq:loglikelihood}
\begin{aligned}
    L = \ln(P(\{v_{r}, \epsilon\}|\mu, \sigma)) = \\
    -\frac{1}{2} \sum^N_{i=1} \left( \ln{(2 \pi(\sigma^2 + \epsilon_i^2))} + 
    \frac{(v_{i,r}-\mu)^2}{\sigma^2 + \epsilon_i^2} \right).
\end{aligned}
\end{equation}

\noindent This expression is the standard Gaussian distribution, with the observational velocity errors taken into account, giving the term $\sigma^2 + \epsilon^2_i$ instead of $\sigma^2$.
This maximization is done using Nelder-Mead optimisation \citep{Gao12_Nelder-Mead} for $\mu$ and $\sigma$ simultaneously.

The error on the derived mean $\mu$ and standard deviation $\sigma$ can be computed by taking the square root of the diagonals of the covariance matrix $\Sigma$, which is related to the Hessian matrix $H$ as $\Sigma = (-H)^{-1}$, where an element in the Hessian is defined as:

\begin{equation}\label{eq:loglikelihooderror}
    H_{i,j} = \frac{\partial^2\log{L}}{\partial\phi_i\partial\phi_j}
\end{equation}

\noindent where $\phi = (\mu, \sigma)$ are the model parameters.

\subsection{Velocity distribution as a function of position}\label{sec:velocitydistribution}

\begin{figure}
 \includegraphics[width=\columnwidth]{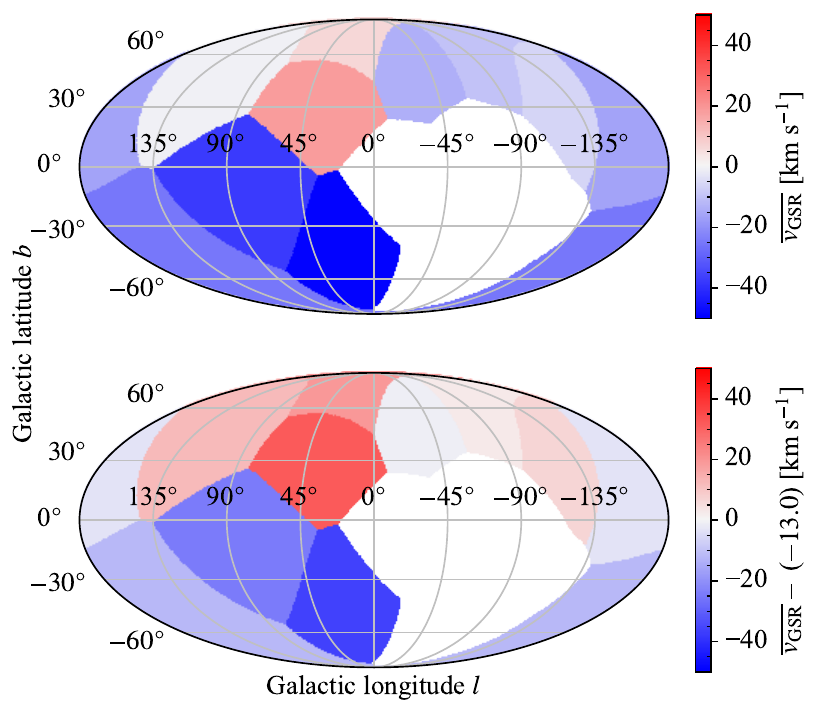}
 \caption{ 
 \textit{Top row:}
 The on-sky distribution of $v_\text{GSR}$ in Galactic coordinates, separated into Voronoi bins, for all stars past 50 kpc. 
 The Southern hemisphere is clearly negative overall, while the Northern hemisphere is consistent with 0 \kms. 
 The errors in each velocity bin range from 7 to 18 \kms.
 \textit{Bottom row:}
 Same as top row, but with the mean velocity of all stars with Galactocentric distance larger than 50 kpc in both hemisphere subtracted.
 This all-sky mean velocity of $-13.0 \pm 3.6$ \kms\ subtracted from the value in each Voronoi bin shows the contrast between the Northern and Southern hemispheres more clearly; the former shows positive velocities and the latter negative velocities.
 } 
 \label{fig:onskyvelocitydistribution}
 \end{figure}

Using the MLE approach, we look into the velocity distribution projected on the sky of the outer halo.
From now on, we will define, and use, the outer halo BHB sample as stars with Galactocentric distances $R>50$ kpc.
To deal with density variations of the sky-projected BHBs and avoid having bins with vastly different numbers of stars, we choose an adaptive binning technique based on Voronoi tessellation on the sphere (Koposov priv. comm.).
The tessellation is built iteratively and aims to have pixels with round shape and close to 85 stars per pixel.
See \citet{Cappellari03_VoronoiSNRbinning, Cappellari09_Voronoiintro} for the implementation of a similar technique for integral field unit data.
For each cell we compute the mean radial velocity and the associated error of stars in the bin, using MLE approach (i.e. Eqs. \eqref{eq:loglikelihood}) and \eqref{eq:loglikelihooderror}). 
For all bins, the error on the mean ranges from 7 to 18 \kms\ and the median distance of stars ranges from 55 to 68 kpc, with most bins having a median distance around $\sim 60$ kpc.
The resulting velocity map can be seen in Fig. \ref{fig:onskyvelocitydistribution}.

In the upper panel of this figure, we see that the entire Southern Galactic hemisphere shows negative velocities.
The Northern Galactic hemisphere has more positive velocities than the Southern, especially for $l > 0^\circ$. 
Across all bins, the mean $v_\text{GSR}$ in the Southern hemisphere is $-34.1 \pm 7.1$ \kms\ and in the Northern hemisphere it is $-3.6 \pm 4.0$ \kms. 
We will refer to these as the bulk velocities of the Southern and Northern hemisphere, respectively.
These bulk velocities are significantly distinct from each other at 3.7$\sigma$.
That the Southern hemisphere has such a strongly negative radial velocity is reflected in Fig. \ref{fig:onskyvelocitydistribution}, where all areas with $b < 0^\circ$ are blue (negative). 
For the Northern hemisphere, the mean being consistent with 0 \kms\ is also reflected in this figure, where the areas with $b > 0^\circ$ are either blue (negative) or red (positive).
The bottom panel shows the same plot as the upper panel, but with the all-sky mean (i.e. across all bins on the entire sky) of --13.0 \kms\ subtracted to emphasize the difference in behaviour between the Northern and Southern hemispheres. 
We can clearly see here that there is a systematic difference in velocities between the two hemispheres, with the Southern hemisphere being much more blueshifted than the Northern hemisphere.

\begin{figure}
 \includegraphics{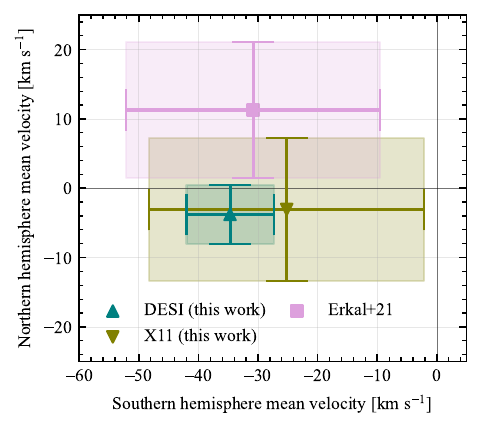}
 \caption{
 Comparison of the bulk velocities of the outer halo in the Southern hemisphere (along x-axis) and the Northern hemisphere (along y-axis) for the values by \citet{Erkal2021LMCouterhalobulkvelocity} (pink), \citetalias{Xue11_SDSSDR8BHBs} (olive) and DESI (teal). 
 } 
 \label{fig:bulkvelocities}
 \end{figure}

In Fig. \ref{fig:bulkvelocities}, the bulk velocities presented above are split into the contributions from the DESI and \citetalias{Xue11_SDSSDR8BHBs} datasets, and compared with the bulk velocities computed by \citet{Erkal2021LMCouterhalobulkvelocity} using BHBs, who found the Southern hemisphere to have a bulk velocity of $-30.8 \pm 21.3$ \kms\ and the North $11.3 \pm 9.8$ \kms.
As the plot shows, for all three data sets, the computed bulk velocities are in agreement both in the Southern and the Northern hemispheres.
The error bar on the mean from the \citetalias{Xue11_SDSSDR8BHBs} sample is much larger than the DESI sample because DESI has more stars past 50 kpc than \citetalias{Xue11_SDSSDR8BHBs}, and so the DESI sample is the one contributing the most to the bulk velocity measurement.

\subsection{Parametrizing the halo velocity field}

Thus far we have only considered the averaged velocity in bins of position, but have not tried to detect the explicit dependence of radial velocity on sky position. 
To understand this, we need to construct a model of the halo's velocity field that describes the observed velocity distributions.
We assume that the velocity field of the halo contains a dipole contribution. 
This is caused by the centre of the MW traveling towards a past point in the LMC's orbit below the Galactic plane \citep{Garavito-Camargo19_LMCinteractionmodels}, which from our point of view causes distant halo stars to become redshifted in the Northern hemisphere and blueshifted in the Southern hemisphere.
The MW's direction of travel, the apex direction, must be included in this description as the contribution from the dipole effect on the velocity field should be maximal at the apex direction, and there should be no contribution from the dipole effect 90$^\circ$ away from the apex direction.
However this dipole contribution is not enough to explain the difference in amplitude in the bulk velocities seen in Fig. \ref{fig:bulkvelocities}, where the amplitude in the North is smaller than in the South.
If the velocity field was to consist of a pure dipole signal, the amplitude in the two hemispheres would be the same.
For the amplitude to be smaller in the North than the South, where the overall velocity is more negative, there must be a negative velocity component acting on the halo across the celestial sphere on top of the velocity dipole, essentially reducing the velocity in the North and making it even more negative in the South.
Since this component is independent of position on the sky and negative, it is globally directed inwards, implying that the halo is being compressed across the entire sky. 
For this reason, we include a monopole term in the model that is directed inwards to the Galactic centre, similar to the mean radial velocity in \citepalias{Petersen21_LMCreflexmotionNature, yaaqib24_Rashid}, which is not dependent on sky position.
Because its existence implies a compression of the halo, we will refer to this velocity as the compression velocity.
This inwards directed and thus negative radial velocity will make the negative dipole velocity in the Southern hemisphere more negative, and it could cancel out the positive dipole velocity in the Northern hemisphere, causing the Northern bulk velocity to be consistent with 0 \kms.
We note that this monopole parameter is introduced to explain the difference in amplitudes between the hemispheres without fully understanding its physical origin, which we discuss in Sec. \ref{sec:discussioncomprvel}.

Combining the dipole and compression contributions, $v_\text{dipole}$ and $v_\text{compr}$ respectively, to the global velocity field, $v_\text{field}$, we can define it as: 

\begin{equation}\label{eq:vglobal}
    v_\text{field} = v_\text{compr} + v_\text{dipole}\cdot \cos{(\theta)}
\end{equation}

\noindent where $\theta$ is the angular distance between the position of a given star $i$ in Galactic coordinates ($l_i, b_i$) and the apex direction ($l_\text{apex}, b_\text{apex}$)\footnote{\url{https://docs.astropy.org/en/stable/api/astropy.coordinates.angular_separation.html} was used for computing $\theta$.}.

We can find $v_\text{compr}$, $v_\text{dipole}$, $l_\text{apex}$ and $b_\text{apex}$ by once again using MLE: we want to find the four parameter values (and a dispersion $\sigma$) that maximizes the likelihood of observing the velocities of our outer halo BHB sample. 
We do this by replacing $\mu$ in Eq. \eqref{eq:loglikelihood} with $v_\text{field}$, keeping the observed velocities $v_r$ and their corresponding errors $\epsilon_i$ the same as in the previous numerical maximization.
Instead of using data in Voronoi cells on the sky as we did previously, we now fit the entire data set.
Because of the geometry of the sphere, we require that $b_\text{apex}$ lies in the range $-90^\circ$ to $90^\circ$ as allowing the latitude to extend past these values is unphysical, but impose no bounds on the other parameters to remain agnostic.
The parameter errors are again computed using Eq. \eqref{eq:loglikelihooderror}.\footnote{
The code used for performing the fit of Eq. \eqref{eq:vglobal} and a tutorial on how to use it is given in \url{https://github.com/abystrom/apex-likelihood-fitting}.}

\begin{figure}
 \includegraphics[width=\columnwidth]{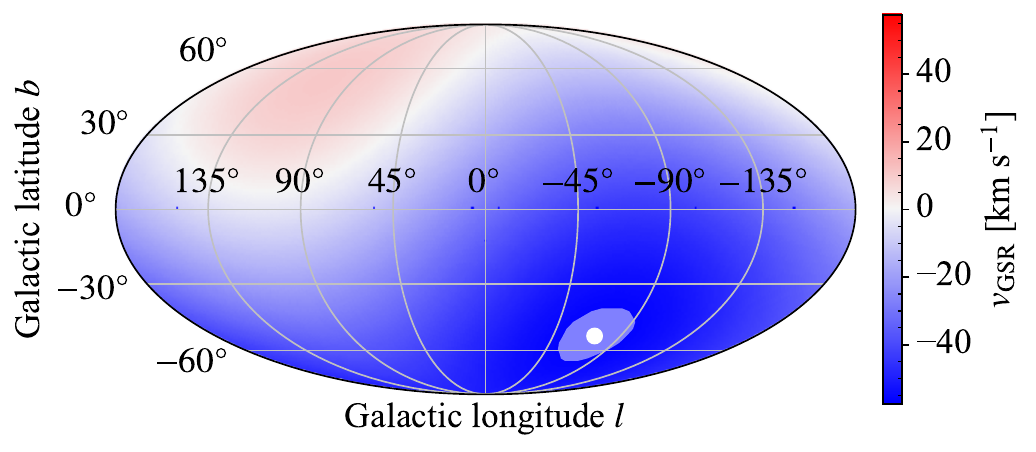}
 \caption{
 The best-fitting on-sky velocity distribution of $v_\text{GSR}$ as computed by applying Eq. \eqref{eq:vglobal} to the BHB sample. 
 The apex direction, from Table \ref{tab:apexfittingresult}, is shown in white.
 } 
 \label{fig:velocityfieldfit}
 \end{figure}

Performing the maximization gives us the parameter values presented in Table \ref{tab:apexfittingresult}.
These values show that globally, the compression velocity $v_\text{compr}$ is directed inwards, and towards the apex direction ($l_\text{apex}$, $b_\text{apex}$), the apex velocity $v_\text{dipole}$ is negative with a higher amplitude than for $v_\text{compr}$.
We note that even though we expect $v_\text{compr}$ to be negative before performing the maximization routine, we do not force the parameter to be negative, and so the value presented in Table \ref{tab:apexfittingresult} is indeed the best-fitting value to our data.
We see the velocity prediction from Eq. \eqref{eq:vglobal} in Fig. \ref{fig:velocityfieldfit}, with the apex direction of the fit shown as a white point. 

\begin{table}
	\centering
	\caption{The results from numerically fitting Eq. \eqref{eq:vglobal} to the BHB sample using MLE. 
                }
        \label{tab:apexfittingresult}
	\begin{tabular}{lr} 
		\hline
		$v_\text{compr}$ [\kms] & $-23.5 \pm 4.6$ \\ 
            $v_\text{dipole}$ [\kms] & $-34.0 \pm 10.6$ \\
            $l_\text{apex}$ [deg] & $-72.7 \pm 21.2$ \\
            $b_\text{apex}$ [deg] & $-52.9 \pm 12.6$ \\
            $\sigma$ [\kms] & $90.5 \pm 2.5$ \\
		\hline
	\end{tabular}
\end{table}

\begin{figure}
 \includegraphics[width=\columnwidth]{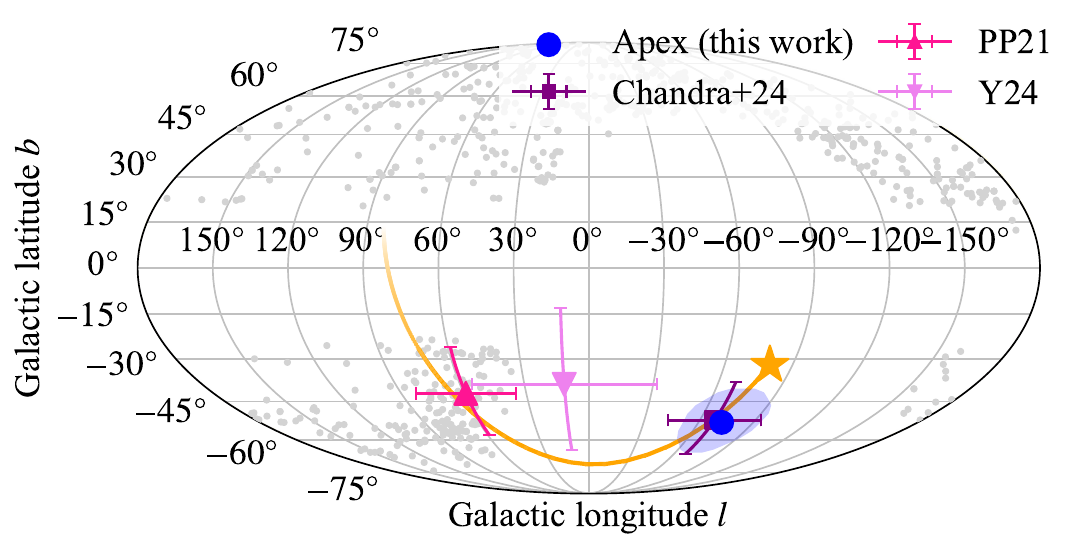}
 \caption{ 
 The apex direction longitude and latitude ($l_\text{apex}, b_\text{apex}$) from Table \ref{tab:apexfittingresult} shown on-sky in blue, and the apex directions from \citet{Chandra24_LMCMW}, \citetalias{Petersen21_LMCreflexmotionNature} and \citetalias{yaaqib24_Rashid} as a purple square, magenta upwards triangle and a violet downwards triangle respectively. 
 For \citetalias{Petersen21_LMCreflexmotionNature} we use their BHB-derived value (using Galactocentric distances larger than 40 kpc), and for \citetalias{yaaqib24_Rashid}, we use their BHB-derived apex direction for the distance bin with Galactocentric distances larger than 50 kpc.
 The distribution of the BHB sample is shown in light gray. 
 The position of the LMC is shown as an orange star with its past trajectory in orange \citep{Vasiliev23_LMChaloreview}. 
        } 
 \label{fig:apexdirections}
 \end{figure}

The parameters presented in Table \ref{tab:apexfittingresult} are similar to those in the halo velocity model by \citetalias{Petersen21_LMCreflexmotionNature}, which is also used by \citetalias{yaaqib24_Rashid} and \citet{Chandra24_LMCMW}:
our $v_\text{compr}$ corresponds to their mean halo radial velocity $\langle v_r \rangle$; $v_\text{dipole}$ to their MW disc barycentre velocity $v_\text{travel}$; and our $(l_\text{apex}, b_\text{apex})$ to theirs. 
The model they use constrains the rotation of the halo by including transverse velocities, while we only use radial velocities in our velocity field parametrization.
These radial velocities are measured from the position of the Sun, and due to the Sun not being at the centre of the Galaxy, our radial velocities will show the dipole and some contribution from the halo's rotation projected onto our line-of-sight.
However we show in Appendix \ref{sec:appendixrotationbias} that for Galactocentric distances larger than 20 kpc the rotation of the halo has a negligible contribution to these radial velocities, and so we can compare their parameters to ours\footnote{\citetalias{Petersen21_LMCreflexmotionNature} and \citetalias{yaaqib24_Rashid} use several different tracer populations, but we only compare to the results using their BHB sample.}.
We see that the velocities as well as $b_\text{apex}$ are consistent between all papers using BHB stars as tracers.
However, our $l_\text{apex}$ does not agree with either work; this means that our velocity field has a different direction than their studies.
The resulting angular offsets between the apex directions are $76^\circ$ and $57^\circ$, respectively. 
This substantial difference means that the velocity field models predict different velocities across the sky.
Our apex direction ($l_\text{apex}, b_\text{apex}$) is seen on-sky in Fig. \ref{fig:apexdirections} in blue together with the BHB sample in light grey, and the apex directions found by \citetalias{yaaqib24_Rashid}, \citetalias{Petersen21_LMCreflexmotionNature} and \citet{Chandra24_LMCMW} in pink, magenta and purple, respectively, with the position of the LMC as an orange star with its past orbit also in orange.
Note that while this work and \citetalias{yaaqib24_Rashid} use $R>50$ kpc, the works by \citetalias{Petersen21_LMCreflexmotionNature}, \citet{Chandra24_LMCMW} use $R>40$ kpc.

Converting the apex directions to Magellanic Stream coordinates ($L_\text{MS}$, $B_\text{MS}$) \citep{Nidever08_MScoordinates} we find that our apex direction is at ($L_\text{MS, apex}$, $B_\text{MS, apex}$) = ($-20.8 \pm 13.2$, $0.7 \pm 15.5$) degrees. 
Even taking the error bar on $L_\text{MS, apex}$ into account, our apex direction is only consistent with trailing behind the LMC by about 50 Myr, confirming what is visually seen in Fig. \ref{fig:apexdirections}, i.e. that the MW disc is moving towards a past point in the LMC orbit.
Compared to \citetalias{Petersen21_LMCreflexmotionNature} and \citetalias{yaaqib24_Rashid} we thus predict a much faster response of the inner MW to the LMC as their apexes are much further along a past point in the LMC orbit, see Fig. \ref{fig:apexdirections}.

Expressing our apex direction in Magellanic stream coordinates lets us directly compare it to the apex direction measured by \citet{Chandra24_LMCMW}: they measure ($L_\text{MS, apex}$, $B_\text{MS, apex}$) = ($-59^{+20}_{-19}$, $25^{+19}_{-18}$) degrees at 40 kpc, and ($L_\text{MS, apex}$, $B_\text{MS, apex}$) = ($-23^{+15}_{-15}$, $-2^{+13}_{-14}$) degrees at 120 kpc, and assume a linear variation of ($L_\text{MS, apex}$, $B_\text{MS, apex}$) in between.
There is perfect agreement between our apex directions, which is seen when comparing the blue dot for our apex direction with the purple cross for their direction in Fig. \ref{fig:apexdirections}.
This is especially encouraging considering the differences in sky coverage between the data used in their work and ours; their work uses data covering the entire sky, whereas ours only covers the DESI footprint, which primarily is missing data in the Southern hemisphere (see Fig. \ref{fig:onskydistribution}), and because we use BHBs as halo tracers and they instead use RGBs.

\section{Model comparison}\label{sec:modelcomparison}

We want to understand if the presented bulk velocities and sky-projected velocity patterns agree with models found in the literature.
To test this, we use the models presented in \citet{Vasiliev24_LMCMWmodels}.
The model by \citet{Erkal20_LMCreflexMWmassbias} is also available, but we choose to work with the former because it contains six models instead of only one.
\citet{Vasiliev24_LMCMWmodels} aims to test if the LMC is on its second pericentric passage, investigating the consensus that it just passed its pericentre for the first time. 
However for both passage scenarios, the radial velocity signal is roughly the same because the most recent passage is the one that mainly dictates the induced kinematic patterns, not the accumulation of all past passages, so we use these models without making any assumption on what point in its orbit around the MW the LMC is currently.
The paper investigates six different models, with two different MW masses (M10 and M11, corresponding to MW masses of $10 \times 10^{11}$ and $11 \times 10^{11}$ M$_\odot$ respectively) and two different LMC masses (L2 and L3, corresponding to LMC masses of $2 \times 10^{11}$ and $3 \times 10^{11}$ M$_\odot$ respectively) creating four combinations (L2M10, L2M11, L3M10, L3M11), with a fifth model assuming that the LMC is on its first pericentric passage (L2M10first) and a sixth assuming a radially anisotropic MW halo (L3M10rad).

When making our comparisons between these models and the data, we apply the DESI footprint as seen in grey in Fig. \ref{fig:onskydistribution} to the models to mimic the DESI footprint\footnote{A tutorial on how to do this footprint selection as well as the DESI footprint file can be found at: \url{https://github.com/abystrom/apex-likelihood-fitting}.}.
The DESI footprint and the \citetalias{Xue11_SDSSDR8BHBs} footprint, with the latter being SDSS data, are almost identical, and so we assume that the DESI footprint is representative of the entire combined BHB sample.

\begin{figure}
 \includegraphics{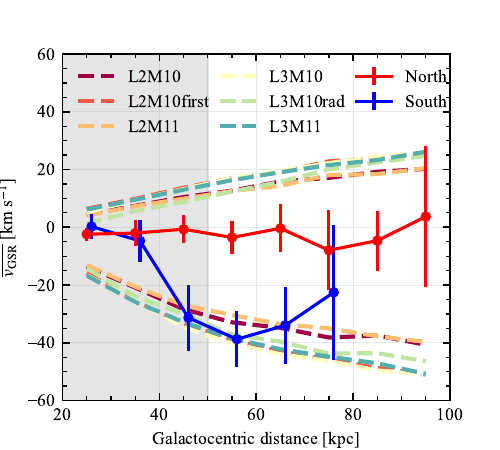}
 \caption{
 Mean $v_\text{GSR}$ in equidistant bins of 10 kpc as a function of distance, for the BHB sample (points with error bars in red for the Northern hemisphere and in blue for the Southern hemisphere) and for all six models from \citet{Vasiliev24_LMCMWmodels} (dashed lines). 
 For the data we require a minimum of 10 stars in each bin.
 L2M11 is the best-fitting model, see Table \ref{tab:modelschisq}.
 Note that the data points in the Southern hemisphere have been shifted by one kpc to the right for increased readability.
 } 
 \label{fig:allmodelsfunctionofdistance}
 \end{figure}

We first wish to understand if the BHB $v_\text{GSR}$ distribution as a function of distance behaves similarly to these models.
This is done by computing the mean $v_\text{GSR}$ and its error in 10 kpc wide Galactocentric radial bins, using Eqs. \eqref{eq:loglikelihood} and \eqref{eq:loglikelihooderror}, for the Northern and Southern hemispheres separately, for both the BHB sample and the halo stellar particles in all of the \citet{Vasiliev24_LMCMWmodels} models.
The comparison between the BHB sample and the models is shown in Fig. \ref{fig:allmodelsfunctionofdistance}.
For Galactocentric distances below 20 kpc, the halo rotation significantly biases fits of Eq. \eqref{eq:vglobal}, as shown in Appendix \ref{sec:appendixrotationbias}.
This means that the halo rotation is large at those distances.
For this reason we only show Galactocentric distances above 20 kpc in the figure, even though the BHB sample extends down to 6 kpc.
In the figure, the mean $v_\text{GSR}$ for the BHB sample is shown as connected points with corresponding error bars in red and blue for the Northern and Southern hemispheres respectively. 
The same for the models are shown as dashed lines.
Looking at this figure, we see that all models have the same qualitative behaviour in velocity as a function of distance, and they all fail to properly fit the data.
In Table \ref{tab:modelschisq}, we compare the reduced $\chi^2$ of each model to the data by comparing the bins in Fig. \ref{fig:allmodelsfunctionofdistance}, for Galactocentric distances above 50 kpc. 
This comparison shows that the L2M11 model is the best fit, and for this reason alone, we will use it for comparison to our data throughout the rest of the paper.

\begin{table}
	\centering
	\caption{
                The reduced $\chi^2$ of each model by \citet{Vasiliev24_LMCMWmodels} compared to the BHB sample, binned as in Fig. \ref{fig:allmodelsfunctionofdistance}.
                }
        \label{tab:modelschisq}
	\begin{tabular}{lc} 
            \hline
            Model & $\chi_\text{red.}^2$ \\
		\hline
            L2M10      & 2.7 \\ 
            L2M10first & 4.1 \\ 
            L2M11      & 2.6 \\ 
            L3M10      & 4.2 \\ 
            L3M10rad   & 3.0 \\ 
            L3M11      & 3.9 \\ 
		\hline
	\end{tabular}
\end{table}

\begin{figure}
 \includegraphics{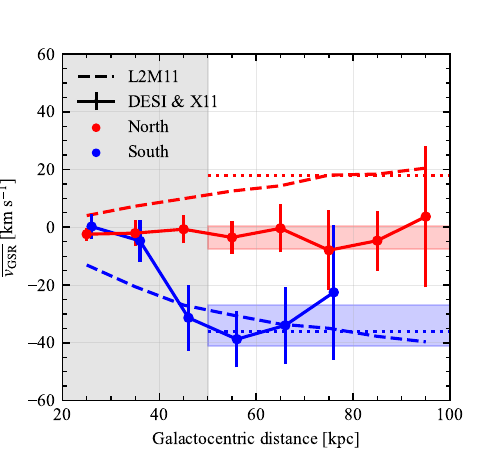}
 \caption{ 
 Mean $v_\text{GSR}$ in equidistant bins of 10 kpc as a function of distance for the data (points with error bars) and for the best-fitting model L2M11 (dashed lines) from \citet{Vasiliev24_LMCMWmodels}, separated into the Northern and Southern hemispheres (red and blue, respectively). 
 For the data we require a minimum of 10 stars in each bin.
 The bulk velocity past 50 kpc for the data, including the error, is seen as shaded rectangles, and the corresponding bulk velocity for the model is seen as dotted lines. 
 Note that the data points in the Southern hemisphere have been shifted by one kpc to the right for increased readability.
 } 
 \label{fig:modelsdatafunctionofdistance}
 \end{figure}

We repeat Fig. \ref{fig:allmodelsfunctionofdistance}, but show only the L2M11 model and the BHB data, in Fig. \ref{fig:modelsdatafunctionofdistance}, and now also include the bulk velocities of all observed and simulated stars with Galactocentric distances exceeding 50 kpc, as rectangles and dotted lines respectively.
This figure shows that as distance increases, both the model and the data show larger radial velocity amplitude for the South, meaning a stronger effect of the LMC on the halo with distance.
The BHB's Northern hemisphere mean velocity on the other hand stays roughly constant with distance, which can also be seen in the models in \citet[their Fig. 4]{Chandra24_LMCMW}.
The higher the distance, the more the Northern and Southern hemispheres deviate in mean $v_\text{GSR}$.
For the BHB sample, the Northern and Southern hemispheres become separated in mean $v_\text{GSR}$ past 40 kpc.
The model instead predicts a much more gradual velocity difference between the Northern and Southern hemispheres, as they are separated already at small Galactocentric radii.
In the Southern hemisphere, the bulk velocity of both the BHB sample and the model agree, but that is not the case for the Northern hemisphere, as the BHB sample's bulk velocity is significantly lower than the model's. 
This can also be seen in the red data points, which are consistently below the red dashed line.

\begin{figure}
 \includegraphics[width=\columnwidth]{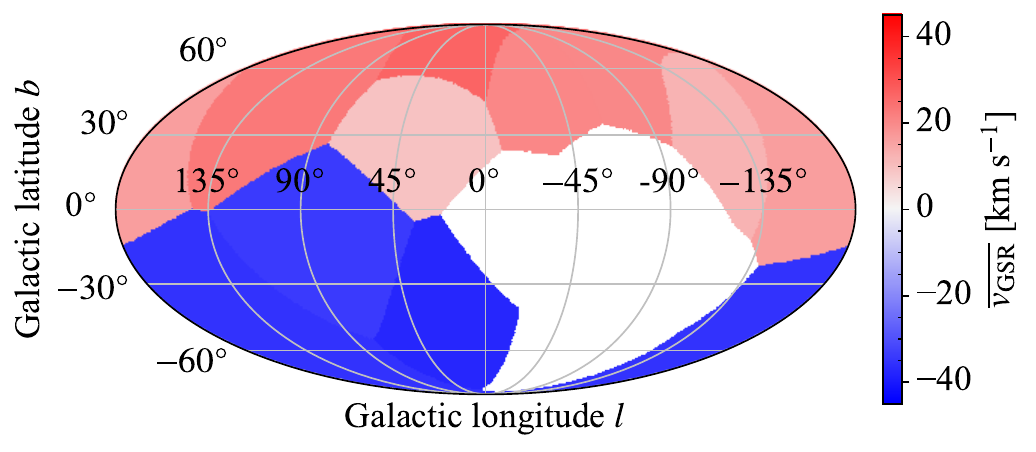}
 \caption{
 The on-sky distribution of $v_\text{GSR}$ of the L2M11 model from \citet{Vasiliev24_LMCMWmodels} in Galactic coordinates, separated into the same Voronoi bins used in Fig. \ref{fig:onskyvelocitydistribution} (with the same empty bins as in that plot masked out in white), for all stars with Galactocentric distance between 50 and 120 kpc, with the on-sky selection representative of the DESI footprint given in Eq. \ref{eq:DESIfootprint}. 
 } 
 \label{fig:onskyvelocitydistributionmodel}
 \end{figure}

If we again look at all the stars in the best-fitting model L2M11 that have Galactocentric distances between 50 and 120 kpc (where the upper range reflects the distance distribution of our sample, see Fig. \ref{fig:distancedistributions}), and use the same Voronoi bins as in Fig. \ref{fig:onskyvelocitydistribution}, we can construct its on-sky velocity distribution.
This is seen in Fig. \ref{fig:onskyvelocitydistributionmodel}, where the bins that are empty in the version for the BHB are also empty for the model because we apply the DESI footprint.

The on-sky velocity distribution of the L2M11 model looks like the distribution of our BHB data in the Southern hemisphere, seen in Fig. \ref{fig:onskyvelocitydistribution}.
The Northern hemisphere looks much more different: the entire Northern hemisphere is coloured red i.e. is redshifted in the model, where the data instead shows both blue and red bins.
This is also reflected in Figs. \ref{fig:bulkvelocities} and \ref{fig:modelsdatafunctionofdistance}, which shows that the data set's Northern bulk velocity is consistent with 0 \kms, showing up as both red and blue spots in Fig. \ref{fig:onskyvelocitydistribution}, whereas the bulk velocity of the model has a larger amplitude.
This disagreement in the Northern hemisphere between the model and the data can be interpreted as a much higher compression velocity in the data. 

To understand if the discrepancy between the model and the data really is due to differences in compression velocity, we apply the velocity decomposition of Eq. \eqref{eq:vglobal} to the L2M11 model, over the distance range 0--120 kpc and with the DESI footprint applied.
Doing so gives us the values in Table \ref{tab:apexfittingresultmodel}, which comparing to the corresponding values for the BHB sample shown in Table \ref{tab:apexfittingresult} shows that the data gives a $v_\text{compr}$ that is $\sim 10$ \kms\ larger in amplitude and a $v_\text{dipole}$ that is $\sim 10$ \kms\ smaller in amplitude than the model.
This means that the model's velocity field more purely consists of a dipole than the data's velocity field.
This explains why the model and the data agree in the Southern hemisphere in Figs. \ref{fig:modelsdatafunctionofdistance} and \ref{fig:allmodelsfunctionofdistance}, where the two velocity components are added, and the $\sim 20$ \kms\ difference between model and data in the North.
We also note that the apex direction for L2M11 agrees much better with our work than it does for \citetalias{Petersen21_LMCreflexmotionNature}, \citetalias{yaaqib24_Rashid}.
The angular offset in apex direction compared to our results differ by about 30$^\circ$, and it differs from the directions by \citetalias{Petersen21_LMCreflexmotionNature} and \citetalias{yaaqib24_Rashid} by 47$^\circ$ and 45$^\circ$ respectively.
This angular offset to \citetalias{Petersen21_LMCreflexmotionNature} was pointed out already by \citet[their Sec. 3.2]{Vasiliev24_LMCMWmodels}, when comparing the sky-projected velocity fields.

\begin{table}
	\centering
	\caption{The results from fitting Eq. \eqref{eq:vglobal} to L2M11 which is the best-fitting model from \citet{Vasiliev24_LMCMWmodels} to our data using MLE.}
        \label{tab:apexfittingresultmodel}
	\begin{tabular}{lr} 
		\hline
		$v_\text{compr}$ [\kms] & $-11.2$ \\ 
            $v_\text{dipole}$ [\kms] & $-40.5$ \\
            $l_\text{apex}$ [deg] & $-25.3$ \\
            $b_\text{apex}$ [deg] & $-78.9$ \\
            $\sigma$ [\kms] & $98.3$ \\
		\hline
	\end{tabular}
\end{table}

\section{Discussion}\label{sec:discussion}

In this section we discuss possible limitations of our study and comparison with literature measurements and implications of our results.
We split the discussion section into two main parts.
In Sec. \ref{sec:discussion_BHBsample} we discuss limitations of our sample such as the metallicity dependence on the BHB distances and how substructure still left in the sample could affect our results.
In Sec. \ref{sec:discussioncomprvel} we discuss the compression velocity parameter $v_\text{compr}$ from Eq. \eqref{eq:vglobal} and in Sec. \ref{sec:compareapexdirectionliterature} we compare our apex direction ($l_\text{apex}$, $b_\text{apex}$) to those found in the literature.

\subsection{The BHB sample}\label{sec:discussion_BHBsample}

\subsubsection{Metallicity dependence of BHB distances}

When computing the absolute magnitudes of the DESI BHBs using Eq. \eqref{eq:Mgcolourrelation}, the only parameter in the function is the $(g-r)$ colour.
However, BHB luminosities are affected by their metallicity; a more metal-poor BHB star will be systematically brighter than a more metal-rich BHB star of the same colour \citep{Fermani13_BHBmetallicityanddistance}.
This is an effect that has been ignored in this work and we must make sure that it does not affect our results. 
The distances to the \citetalias{Xue11_SDSSDR8BHBs} BHBs that we use in this work are the same as in the original catalogue, and were not derived using Eq. \eqref{eq:Mgcolourrelation} so we are only interested in testing the DESI BHB distances for metallicity dependence.
If the metallicity gradient with distance for BHBs is significant one could imagine having a systematic effect on our results because of our distance binning.

We need to make sure that the BHB distance dependence on metallicities' are not affecting the results presented in this paper.
We do this by ensuring that the bulk velocities of the Northern and Southern hemispheres are the same for the metal-poor and the metal-rich BHBs in the sample.
For this test we only use BHBs with metallicities in the range [-3.0, 0.0] dex, whereas the full sample used for the scientific analysis contains stars both more metal-poor and metal-rich.
This way we exclude stars close to the grid edges as those metallicities are less reliable.
We split our sample of BHB stars into a metal-poor sample ([Fe/H] $\in [-3.0, -1.5]$) and a metal-rich sample ([Fe/H] $\in [-1.5, 0.0]$), and compute the bulk velocity of the Northern and Southern hemispheres for both bins.
The resulting bulk velocities are shown in Table \ref{tab:metallicitydistancebulkvelocity}, and they agree with each other within the error bars.
Fitting the model defined by Eq. \eqref{eq:vglobal} to stars in these metallicity bins also yields parameters that agree with each other within the error bars. We therefore conclude that the theoretical metallicity dependence of the BHB distances does not affect the results in this work.

\begin{table}
	\centering
	\caption{
                 A comparison of the bulk velocity of the entire DESI BHB sample for the Northern and Southern hemispheres, to the same values for the sample sliced into a metal-poor and a metal-rich bin, to test if the metallicity dependence on BHB distances ignored in Eq. \eqref{eq:Mgcolourrelation} affects the velocity distributions studied in this work.
                 The velocities and their errors are computed by minimizing Eq. \eqref{eq:loglikelihood} and applying Eq. \eqref{eq:loglikelihooderror}.
                 Because the values for each hemisphere agrees within the error bars, we conclude that there is no bias from metallicities on the results in this work.
 }
	\label{tab:metallicitydistancebulkvelocity}
	\begin{tabular}{cccc} 
		\hline
		\text{} [Fe/H] & $N$ & $v_{\text{GSR}, b>0^\circ}$ [\kms] & $v_{\text{GSR}, b<0^\circ}$ [\kms] \\
		\hline
        All & 617 & $-3.8 \pm 4.3$  & $-34.7 \pm$ 7.3 \\
		\text{} [--3.0, --1.5] & 313 & $-1.2 \pm 5.8$  & $-28.0 \pm 11.2$ \\
		\text{} [--1.5, 0.0] & 170 & $-10.0 \pm 8.8$  & $-43.3 \pm 12.6$  \\
		\hline
	\end{tabular}
\end{table}

\subsubsection{Effects from halo substructure}

To measure the velocity field of the stellar halo, it is necessary to remove substructure in the halo that is not phase-mixed, as it would bias the velocity signal.
We explain in Sec. \ref{sec:data_Sgrflagging} what cuts in heliocentric distance and Sgr-coordinates (\Sgrlon, \Sgrlat) we use to remove both Sgr and Cetus-Palca, and show in the right-hand column of Fig. \ref{fig:Sgrcuts} that our cuts remove both streams from the data set.
It is still important to understand the sensitivity of our results to this stream removal.
This sensitivity is tested in two ways.
First we keep Sgr the data set (and because we do not identify Cetus-Palca stars with Galactocentric distances $R>50$ kpc, this comparison only concerns the Sgr selection). 
Second we do a stream removal only in (\Sgrlon, distance) without the $|\tilde{B}| < 15^\circ$ requirement.
For both of these cases, we recompute the bulk velocities for all stars with Galactocentric distances larger than 50 kpc and compare to the results presented in Sec. \ref{sec:velocitydistribution}.

We start with leaving Sgr in the data set.
This gives a bigger sample to work with as now the Sgr stars are included among the halo stars.
However it only increases the sample size in the Northern hemisphere, as our adopted Sgr selection method does not identify any stars as Sgr members that both have $R>50$ kpc and are in the Southern hemisphere.
This exercise gives us a more complete smooth halo sample, as we will reintroduce wrongly removed halo stars, but it is less pure, as we also reintroduce halo substructure.
If Sgr is left in the Northern hemisphere sample, the bulk velocity is $\langle v_{\text{GSR}, b>0^\circ} \rangle = 4.9 \pm 2.7$ \kms, which is not in agreement with the bulk velocity when we remove these stars: this shows how important removal of substructure from the smooth halo sample is.
This value is also closer to the value reported in \citet{Erkal2021LMCouterhalobulkvelocity}, where Sgr stars are removed in a similar fashion to our Eq. \eqref{eq:Sgr_cut} but with |\Sgrlat| $< 20^\circ$ and an additional velocity cut.
That the bulk velocity increases when Sgr is kept in the halo sample can be understood from Fig. \ref{fig:Sgrcuts}: in the top left panel, we see that for large enough heliocentric distances, the only stream stars are Sgr stars centered around $\tilde{\Lambda} \approx 70^\circ$, and a majority of these stars have velocities $v_\text{GSR} > 0$ \kms.

Now we only consider the two coordinates \Sgrlon\ and heliocentric distance (i.e. the second part of Eq. \eqref{eq:Sgr_cut}) as our Sgr selection.
Because the stars now need to only pass two requirements instead of three to be labelled as stream stars, more stars will receive this label, and removing them means a smaller halo sample to work with that has higher purity but lower completeness than the sample used in this work.
This changes the bulk velocities to $\langle v_{\text{GSR}, b>0^\circ} \rangle = -2.0 \pm 4.6$ \kms\ and $\langle v_{\text{GSR}, b<0^\circ} \rangle = -33.8 \pm 7.2$ \kms.
These bulk velocities are in agreement with the ones reported in Sec. \ref{sec:velocitydistribution}.
The error bars show a small increase, which is due to the reduced sample size.
This smaller sample is more conservative than the full sample used throughout this paper, and because these bulk velocities are in agreement with the presented values in this paper, our results are not sensitive to the removal of Sgr or Cetus-Palca stream stars.

Despite not finding obvious cold structures in velocity after the removal of Sgr and Cetus-Palca, there is still a possibility that our sample contains low-mass substructure.
Several studies suggest that the outer halo was completely built up by mergers, which has been seen both in observations \citep{Naidu2020haloentiresubstructure} and in simulations \citep{Monachesi19_AURIGA, Font20_ARTEMIS, Wright24_FOGGIE}. 
The resulting post merger debris has a bottom heavy mass distribution where there is more low-mass than high-mass debris that become more dominant in numbers as distance from the Galactic centre increase \citep{Fattahi20_halomergerdebrismassdistribution}.
While we cannot identify obvious remnant cold substructures in the BHB sample, perhaps because such potential distant streams have too low surface brightness \citep{Shipp23_FIREstreamsdetectability}, they can still be identified through clustering in phase space \citep{Helmi99_phasespacedebris, Bonaca21_streamsphasespace}.
Performing a phase space clustering search is beyond the scope of the paper.
However if the number of undetected low-mass stellar merger debris in the BHB sample is large enough, we would not expect that it should bias the average velocity measurements in this work, as the average velocities across all structures should be a good proxy of the bulk velocity of the halo.

\subsection{Understanding the compression velocity}\label{sec:discussioncomprvel}

The compression velocity $v_\text{compr}$ that we have used as a parameter in our model of the halo velocity field in Eq. \eqref{eq:vglobal} is measured to be of the same order as the dipole velocity $v_\text{dipole}$ and is thus a significant contribution to the velocity field. 
We introduced it to understand the difference in velocity field amplitudes between the Northern and Southern hemispheres (see Figs. \ref{fig:onskyvelocitydistribution} and \ref{fig:bulkvelocities}), but though it is physically motivated by previous works \citepalias{Petersen21_LMCreflexmotionNature, yaaqib24_Rashid}, in our work it is simply a monopole term in the description of the velocity field and its true nature remains to be determined. 

\citetalias{yaaqib24_Rashid} suggested that the compression velocity (or their mean radial velocity) is due to the LMC's infall to the MW stripping the LMC of mass, which when deposited in the MW's potential deepens the potential well, causing halo stars to fall into the MW.
This argument is strengthened by their simulations, that show that a heavier LMC makes the mean radial velocity across the entire halo more negative. 
The model only considers the global effects of the LMC on the MW halo, as it treats the LMC as a point mass and thus the local wake from dynamical friction is not included in this explanation.
\citet{Patel24_LMCsatellites} argue that the LMC's presence around the MW make the galaxy's satellite galaxies more radially concentrated with time, especially after the LMC's recent pericentric passage, supporting the idea that the LMC deepens the potential well around the MW.
Though it is beyond the scope of this paper to verify this explanation or to fully understand what the parameter $v_\text{compr}$ is as defined in our model, we wish to offer some additional potential explanations, focusing more on the effects of local interactions between the MW and the LMC. 
These will hopefully be tested in future work.

One potential explanation is that the compression velocity is due to small-scale, not fully phase-mixed substructure in the halo, that bias our sample in certain regions on-sky.
This is motivated by our incomplete sky coverage, where the anisotropy of the halo \citep{Dey23_M31DESI, Amarante24_haloanisotropy} could become important.
These structures would manifest as chevron-like features in $(R, v_\text{GSR}$) space \citep{Bullock05_modelsRVGSR, Belokurov23_dataRVGSR} that we do not see in our data, but we cannot rule out that they are present at low significance.

The region in the Southern hemisphere where we have data may appear to have a high velocity amplitude compared to the Northern hemisphere, creating a larger $v_\text{compr}$, because it overlaps with the local halo overdensity trailing the orbit of the LMC induced by the LMC's infall into the MW \citep{Amarante24_haloanisotropy}, the local wake.
This wake has an even more negative mean radial velocity than induced by the dipole \citep{Garavito-Camargo19_LMCinteractionmodels, Vasiliev23_LMChaloreview} and covers distances between 45 and 100 kpc \citep{Garavito-Camargo19_LMCinteractionmodels}, peaking at 70 kpc \citep{Amarante24_haloanisotropy}, and could thus potentially overlap with our data set.
If it is a significant part of our data set, it would manifest itself as a seemingly large value of the compression velocity because it reduces the overall velocity observed in the Galactic South.

\begin{figure}
 \includegraphics[width=\columnwidth]{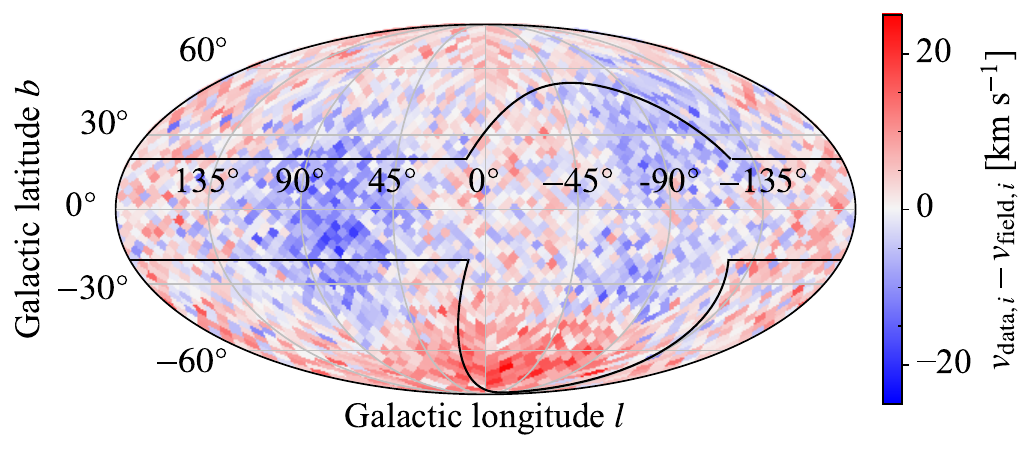}
 \caption{
 The residuals of the stellar particle velocities in the L2M11 model \citep{Vasiliev24_LMCMWmodels} and their predicted velocities from the fit of the parameters of Eq. \eqref{eq:vglobal} in Table \ref{tab:apexfittingresultmodel} shown in HEALPix\protect\footnotemark.
 The DESI observational footprint in Eq. \eqref{eq:DESIfootprint} is shown with black lines.
 This sky map reveals structure in the velocity field residuals.
 The most significant region is the very negative region seen at $(l, b) \approx (75 ^\circ, -15^\circ$), which roughly corresponds to the position of the local wake.
 } 
 \label{fig:L2M10modelvelocityresiduals}
 \end{figure}

Our parametrization of the velocity field does not contain a local wake component, and if it is present in the data but not in the model, the expected velocity field from Eq. \eqref{eq:vglobal} subtracted from the data should reveal a negative spot at the location of the local wake.
We test this by redoing the fit of Eq. \eqref{eq:vglobal} on all models by \citet{Vasiliev24_LMCMWmodels}, and subtract this fitted velocity field from the simulated velocity data.
This is shown for the L2M11 model in Fig. \ref{fig:L2M10modelvelocityresiduals}, that also includes an approximation of the DESI footprint as given by:

\begin{equation}\label{eq:DESIfootprint}
\begin{aligned}
    |b| > 20^\circ \\
    \text{ and } \\
    [(b>0^\circ) \text{ and } (\text{Dec.} > -10^\circ)] \text{ or } [(b<0^\circ) \text{ and } (\text{Dec} > -30^\circ)].
\end{aligned}
\end{equation}

\noindent This reveals a region with velocity more negative than the dipole prediction from our model, centered at $l \approx 75 ^\circ$ for all six models, with a center in $b$ varying between $\sim -10^\circ$ -- $0^\circ$.
The amplitude of this region is $\sim 10-20$ \kms.
This range roughly encompasses our measured velocity compression of --23.5 \kms.
We do not see particularly strong negative residuals in our model in the $(l, b) \approx (75 ^\circ, -15^\circ$) region, see Fig. \ref{fig:appendixdatamodelresiduals}, suggesting that the potential presence of the local wake does not affect our data nor the model fit to it.

\footnotetext{\url{https://sourceforge.net/projects/healpix/}}

The Pisces plume stream has been attributed to the local wake \citep{Belokurov19_Piscesplumelocalwake}, and we test the presence of the local wake by reproducing the right panel of Fig. 1 in \citet{Belokurov19_Piscesplumelocalwake} showing (\Sgrlon, distance) for \Sgrlat\ ranging from --10 to 10 degrees and ranging from 10 to 30 degrees, but there is no visible overdensity there. 
Again, we note that the absence of a density signal is not proof that we do not have Pisces plume stars in our data set.
We also note that the potential presence of the local wake would not explain the low amplitude in the Northern hemisphere seen in Fig. \ref{fig:allmodelsfunctionofdistance}, which is lower than expected for the models in \citet{Vasiliev24_LMCMWmodels}.
The low amplitude in the Northern hemisphere is what prompted us to introduce $v_\text{compr}$ as a parameter in our velocity field model, not the high amplitude in the Southern hemisphere, which is in agreement with the models by \citet{Vasiliev24_LMCMWmodels}.

If we repeat the fit of Eq. \eqref{eq:vglobal} to the models L2M10, L2M11, L3M10 and L3M11 by \citet{Vasiliev24_LMCMWmodels}, we can compare the resulting compression velocities and see how they change as a function of MW and LMC mass.
We ignore the models L2M10first and L3M10rad to let the masses be the only variable parameters.
The resulting values are given in Table \ref{tab:vcompressionLMCMWvaryingmasses}.
This shows that as the LMC mass increases, the amplitude of $v_\text{compr}$ increases.
The inverse is true for the MW mass, so that the amplitude of $v_\text{compr}$ decreases as the MW mass increases.
This means that the $v_\text{compr}$ amplitude is maximized for L3M10 and minimized for L2M11.
Investigating this further is beyond the scope of this paper, but to understand why $v_\text{compr}$ changes as a function of LMC and MW mass, dedicated models are required.

\begin{table}
	\centering
	\caption{
        The results from redoing the fits of Eq. \eqref{eq:vglobal} on the compression velocity using the four models by \citet{Vasiliev24_LMCMWmodels} that all have the same potential, but with varying LMC and MW masses.
        This shows that as the LMC mass increases, the compression velocity amplitude increases, and when the MW mass increases, the compression velocity amplitude instead decreases.
        }
        \label{tab:vcompressionLMCMWvaryingmasses}
	\begin{tabular}{cc}
		\hline
		Model & $v_\text{compr}$  \\
		\hline
		L2M10 & $-12.4 \pm 0.2$ \\
		L2M11 & $-11.2 \pm 0.2$ \\
		L3M10 & $-15.5 \pm 0.2$ \\
		L3M11 & $-13.9 \pm 0.2$ \\
		\hline
	\end{tabular}
\end{table}

\subsection{Comparison with literature apex directions}\label{sec:compareapexdirectionliterature}

The BHB sample presented in this work is the largest spectroscopic BHB sample in the literature, and will only grow with upcoming DESI data releases.
The DESI BHB sample contains 9,866 BHBs, and SDSS DR6 contains 2,401 BHBs \citep{Xue08_SDSSBHBSample} and SDSS DR8 contains 4,985 BHB stars \citepalias{Xue11_SDSSDR8BHBs}.
The catalogue's strengths is not only its increased number of BHB stars compared to previous spectroscopic works: the DESI survey also reach fainter magnitudes leading to a larger probed distance range.
It also has smaller velocity errors than previous works.
That we only use BHB stars instead of combining different halo tracers, such as RGB stars, leads to a less complex sample, that is homogeneous across the entire probed distance range.
Comparing the apex direction we derive using our BHB sample to that published in \citet{Chandra24_LMCMW}, we are in agreement.
They use K-giants across the entire sky and we instead have limited sky coverage (see Fig. \ref{fig:onskydistribution}). 
However, if we compare our direction to those published by \citetalias{Petersen21_LMCreflexmotionNature} and \citetalias{yaaqib24_Rashid} using spectroscopic BHB tracers from SDSS, our dipole directions disagree, with an angular offset of $76^\circ$ and $57^\circ$ respectively (as seen in Fig. \ref{fig:apexdirections}).
The aforementioned works utilise Gaia proper motions, while we solely use radial velocities.
Interestingly, the amplitude of the velocity fields agree. 
Because of this, there is rough agreement in predicted halo velocities where $l > 0^\circ$, but not complete agreement across the celestial sphere, and this lack of complete agreement is due to the differences in apex directions.
We must understand where this disagreement comes from: is it due to our different data sets, or because of the different models we use for our velocity fields?

We will first rule out that the disagreement between our results is because of our model defined in Eq. \eqref{eq:vglobal}.
Our model is one-dimensional and incorporates only radial information, while theirs has three dimensions: the radial, and the two angular directions, simultaneously modeling radial velocity and proper motions.
We note that \citet{Chandra24_LMCMW} also use this three-dimensional model, but still compute an apex direction that is completely in agreement with ours.
Still, we must make sure that without proper motions, our model is complex enough to be a good fit to the data.
In appendix \ref{sec:appendixmodelfittodata} we confirm the validity of our model by first looking at the on-sky residual between our measured velocities and those predicted by our model, and see that there is no pattern with residuals on-sky, meaning that all parts of the sky are modelled equally well.
We then compare the distribution of these residuals with a theoretical Gaussian distribution of mean 0 and the standard deviation being the sum of the standard deviations of the observed velocities and predicted velocities which would be the case for no residuals between observed and predicted velocities.
We find that these two distributions are statistically the same.
These two tests show that the model in Eq. \eqref{eq:vglobal} describes our data well.
The final thing to make sure is that our model can still produce the results by \citetalias{Petersen21_LMCreflexmotionNature} and \citetalias{yaaqib24_Rashid} using their data, which it can, meaning that it is not the model leading to disagreeing results. 
We thus conclude that the disagreement does not stem from our model being a bad fit to the data.

The other possible explanation for the tension of results is differences in the BHB samples.
The works by \citetalias{Petersen21_LMCreflexmotionNature} and \citetalias{yaaqib24_Rashid} use K-giant and BHB tracer samples, but we continue to focus on their results derived using BHBs.
Their BHB samples are based on the SDSS sample by \citetalias{Xue11_SDSSDR8BHBs}, which is the same sample we use for the SDSS portion of our sample.
Comparing their BHB sample to the combined one used in this work, we see that their sample is different from ours in distance and spatial distribution for stars with Galactocentric distance larger than 50 kpc.
We can see in Fig. \ref{fig:distancedistributions} that their sample is more concentrated around the Galactic centre and that we have more stars at greater distances: they have 2.2 \pc\ of stars past 50 kpc (7.1 \pc\ past 40 kpc), while we have 7.8 \pc\ past 50 kpc (or 16.1 \pc\ past 40 kpc).
\citetalias{Petersen21_LMCreflexmotionNature} used an outer halo sample with $R > 40$ kpc, where we instead chose to use $R > 50$ kpc.
If we instead use  $R > 40$ kpc and redo the fit of the halo velocity field on the DESI BHB sample, our $(l_\text{apex}, b_\text{apex})$ equals $(-65.2, -52.4)$ degrees, which is closer to the apex direction from \citetalias{Petersen21_LMCreflexmotionNature} than the one for $R > 50$ kpc which we present in this paper, though only by about 2 degrees.
The two samples have roughly the same sky coverage, but their work measures more stars with $l < -100$ degrees, which is the main region in which our predicted velocities differ (and they mostly agree where $l > 0^\circ$), and our work is measuring more stars in the Southern hemisphere ($b < 0^\circ$).
If we subtract the predicted velocity fields from \citetalias{Petersen21_LMCreflexmotionNature} and \citetalias{yaaqib24_Rashid} from ours (Fig. \ref{fig:velocityfieldfit}), we also see that our data (Fig. \ref{fig:onskydistribution}) extends further into regions of larger discrepancy than their data does.  
The differences in distance and spatial distributions between the data sets are thus subtle and difficult to isolate, but we can see from the appendix in \citetalias{yaaqib24_Rashid} that that could still affect the sample.

The BHB data sets could also be different in terms of purity and completeness since they use different spectroscopic information to clean the BHBs from BSs and main sequence stars.
We test this in Appendix \ref{sec:appendixDESISDSScomparison}, where we compare the BHB stars of both samples with the candidate BHB stars of the other. 
This comparison shows us that both data sets are pure, but that the DESI BHB sample is more complete than the SDSS BHB sample, originally presented in \citet{Xue08_SDSSBHBSample}.

The final major difference between our data set and those of \citetalias{Petersen21_LMCreflexmotionNature}, \citetalias{yaaqib24_Rashid} is the masking of Sgr stars.
We compare the two methods of selecting Sgr stars on the DESI BHB stars in Appendix \ref{sec:appendixcompareSgrcuts}.
The comparison shows that the method used by \citetalias{Petersen21_LMCreflexmotionNature}, \citetalias{yaaqib24_Rashid} leaves Sgr stars in our BHB sample as visible overdensities, especially at the Sgr leading arm apocentre, and all of the Cetus-Palca stream (we note that Cetus-Palca is not present in their data and that it is left in our sample does not imply that it is also left in theirs).
The stream stars that are not removed from our BHB sample and at Galactocentric distances larger than 50 kpc, and thus enter into the analysis, have largely positive radial velocities and lie in the Northern hemisphere, see Fig. \ref{fig:onskydistribution} where the Sgr stream track is shown in purple.
The treatment of Sgr stars in our samples thus appears to be a significant source of differences in our apex directions, and the way it is removed from the working sample will systematically bias the point of the apex direction due to the location of the stream in the Northern hemisphere. 

\section{Summary and conclusions}\label{sec:summaryconclusions}

In this work, we have constructed a sample of BHB stars from the DESI DR1 and Y2 spectroscopic observations, all with radial velocities and distances.
This is the largest spectroscopic BHB sample in the literature, that also has the largest number of distant BHBs due to the faint magnitudes that the DESI survey can reach.
The errors on the BHB radial velocities in DESI are $\sim 2.2$ \kms, and are significantly lower than in previous catalogues.
Because of the large distances probed and the precise velocities, this BHB sample is an excellent sample to study the interaction between the MW and the LMC using distant halo stars. 
The catalogue also has several other potential applications, such as constraining the Milky Way mass profile.

In this work the DESI BHBs are combined with BHBs from \citetalias{Xue11_SDSSDR8BHBs} to create a sample of 13,076 spectroscopically confirmed BHB stars, out of which 10,650 are used in the analysis of the smooth stellar halo to investigate the interaction between the MW and the LMC. 
We especially focus on the 675 stars in the outer halo that have Galactocentric distances larger than 50 kpc. 
The majority of these stars, 617, are provided by DESI, demonstrating the power of DESI to probe far into the distant MW halo.

The analysis of the Galactic standard of rest velocities, $v_\text{GSR}$, of these stars reveals that:

\begin{itemize}
    \item 
    The stars in the Southern hemisphere have a negative bulk radial velocity of $-34$ \kms, while stars in the Northern hemisphere have a bulk radial velocity that is consistent with 0 \kms.
    These two bulk velocities are distinct from each other at 3.7$\sigma$ and can be interpreted as the disc moving to the Galactic South towards a past point of the LMC orbit. 
    
    \item 
    We model the outer halo's velocity field projected on the sky by decomposing it into a monopole and a dipole velocity and fit these velocities together with the apex direction, which is the position of the peak of the velocity field, in Eq. \eqref{eq:vglobal}.
    The monopole term is measured to be $-24$ \kms, and has no angular dependence. 
    The dipole velocity is $-34$ \kms and peaks at the so-called apex direction ($l_\text{apex}, b_\text{apex}) = (-73^\circ, -53^\circ$).
    This can be interpreted as the MW disc moving towards a point along the LMC trajectory.
    
    \item 
    Our apex direction is in good agreement with that by \citet{Chandra24_LMCMW} but significantly far away from measurements by \citetalias{Petersen21_LMCreflexmotionNature} and \citetalias{yaaqib24_Rashid}.
    We believe that the latter disagreement is to some degree explained by different methods in removing the Sgr stream stars from the outer halo samples. 
    
    \item 
    The reason we include a monopole term in our velocity field parametrization is because the measured velocities in the North are close to zero but strongly negative in the South, meaning that the velocity field is inconsistent with just a dipole component. 
    As the monopole term is found to be negative, we interpret it as the halo stars globally moving inwards, so we refer to this term as a compression velocity.
    
    \item 
    In the inner halo we observe that the bulk velocities of stars in the two hemispheres are consistent with each other within 40 kpc and start to diverge beyond that.
    This means that the inner ($<40\,$kpc)  halo is comoving with the MW disc.
    
    \item 
    We compare our observations to models by \citet{Vasiliev24_LMCMWmodels}. 
    The comparison shows that the models show some compression velocity, but at about half of the value observed in the MW. 
    The reason for this disagreement is not clear.
    
\end{itemize}

There are still some unanswered questions after this work, such as what is the true nature of the compression velocity.
However, we are clearly seeing the effects that the LMC has on the MW. 
The LMC's perturbation of the MW results in a global velocity field of the outer halo that is blueshifted towards the LMC in the Southern hemisphere, and more redshifted in the Northern hemisphere. 
The LMC sets the MW in disequilibrium, where the inner regions of the galaxy are moving with respect to the outer halo.

To fully explain the MW--LMC interaction, future work should provide a better understanding of the theory behind this interaction.
There are only a handful of simulations in literature that investigate this interaction in enough detail (e.g. \citealp{Gomez15_MWLMCinteractionmodel, Garavito-Camargo19_LMCinteractionmodels, Garavito-Camargo21_LMCBFE, Erkal20_LMCreflexMWmassbias, Petersen20_LMCreflexMWpotentialbias, Lilleengen23_OCstreamLMC, Vasiliev24_LMCMWmodels}) and they differ from each other.
More models are needed specifically to match the existing set of observational constraints, such as the results presented in this work.

While more observational and theoretical work is necessary to understand the details of the MW-LMC interaction, the exquisite DESI data presented in this paper increase the observed number of outer halo stars significantly and distinctly showcase the LMC's effect on the MW disc and halo. 
Given that the existing DESI data covers Northern sky, while LMC is in the South, it is essential however to complement the existing outer halo samples by observations of Southern sky with surveys like S5 \citep{Li19_S5} and the upcoming 4MOST \citep{deJong19_4MOST}. 
In the North however, future DESI data releases based on data beyond the initial two years of observation used in this study, as well as possible DESI extension beyong the nominal five years of observations, will help to establish a much clearer picture of the MW--LMC interaction. 

\section*{Acknowledgements}

AB thanks Mike Petersen and Rashid Yaaqib for insightful discussions.
AB also acknowledges the Edinburgh Doctoral College Scholarship that funded her work.
SK acknowledges support from the Science \& Technology Facilities Council (STFC) grant ST/Y001001/1.
T.S.L. acknowledges financial support from Natural Sciences and Engineering Research Council of Canada (NSERC) through grant RGPIN-2022-04794.

This work made use of the Python packages \texttt{numpy} \citep{harris20_numpy}, \texttt{matplotlib} \citep{Hunter07_matplotlib}, \texttt{scipy} \citep{Virtanen20_scipy}, \texttt{healpy} \citep{Gorski05_HEALPix, Zonca19_healpy}, \texttt{gala} \citep{Price-Whelan17_gala, adrian_price_whelan_2020_4159870}, and \texttt{astropy} \citep{Astropy13_astropy, Astropy18_astropy, Astropy22_astropy}. 
It also made use of \texttt{sqlutilpy} with DOI \url{https://doi.org/10.5281/zenodo.5160118}. 
This paper made use of the Whole Sky Database (WSDB) created and maintained by Sergey Koposov at the Institute of Astronomy, Cambridge with financial support from the Science \& Technology Facilities Council (STFC) and the European Research Council (ERC).

This material is based upon work supported by the U.S. Department of Energy (DOE), Office of Science, Office of High-Energy Physics, under Contract No. DE–AC02–05CH11231, and by the National Energy Research Scientific Computing Center, a DOE Office of Science User Facility under the same contract. Additional support for DESI was provided by the U.S. National Science Foundation (NSF), Division of Astronomical Sciences under Contract No. AST-0950945 to the NSF’s National Optical-Infrared Astronomy Research Laboratory; the Science and Technology Facilities Council of the United Kingdom; the Gordon and Betty Moore Foundation; the Heising-Simons Foundation; the French Alternative Energies and Atomic Energy Commission (CEA); the National Council of Humanities, Science and Technology of Mexico (CONAHCYT); the Ministry of Science and Innovation of Spain (MICINN), and by the DESI Member Institutions: \url{https://www.desi.lbl.gov/collaborating-institutions}. Any opinions, findings, and conclusions or recommendations expressed in this material are those of the author(s) and do not necessarily reflect the views of the U. S. National Science Foundation, the U. S. Department of Energy, or any of the listed funding agencies.

The authors are honored to be permitted to conduct scientific research on Iolkam Du’ag (Kitt Peak), a mountain with particular significance to the Tohono O’odham Nation.

\section*{Data Availability}

All DESI spectroscopically confirmed BHBs (see Sec. \ref{sec:data_BHBselection}) and their distances (see Sec. \ref{sec:data_distances}) that are part of the DESI DR1 dataset are available at: \url{https://data.desi.lbl.gov/doc/releases/dr1/vac/mws-bhb/}.
The DESI Y2 BHB subsample will be made available in DESI DR2.
The data behind all figures in this paper are available as .fits files at: \url{https://doi.org/10.5281/zenodo.13711675}.
The code used to fit the parameters in Eq. \eqref{eq:vglobal} and to choose a DESI-like sky footprint (see Sec. \ref{sec:modelcomparison}) are available at: \url{https://github.com/abystrom/apex-likelihood-fitting/}.


\bibliographystyle{mnras}
\bibliography{bib} 



\appendix

\section{Stars selected as Sgr or Cetus-Palca}\label{sec:appendixSgrselected}

\begin{figure}
 \includegraphics{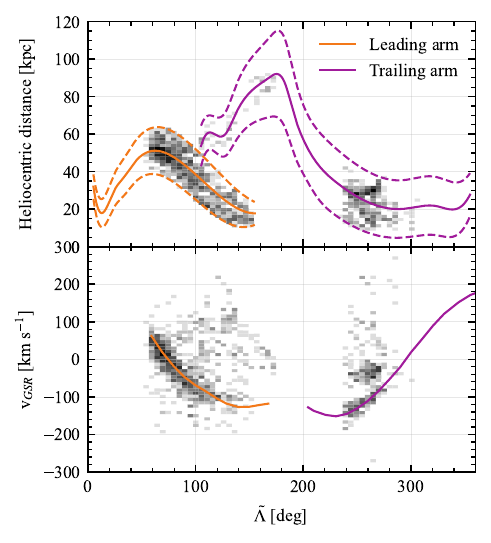}
 \caption{ 
 Same as Fig. \ref{fig:Sgrcuts} but showing the stars selected as Sgr and Cetus-Palca members.
 In the bottom panel, Cetus-Palca clearly stands out as the structure vertically offset from the Sgr velocity track around \Sgrlon\ $\approx 250^\circ$, at $\sim 50$ \kms\ \citep{Yuan19_Cetus}.
 } 
\label{fig:appendixSgrcolumnrownormalised}
\end{figure}

We show the stars selected as belonging to the Sgr or Cetus-Palca streams in (\Sgrlon, distance) and (\Sgrlon, $v_\text{GSR}$) in Fig. \ref{fig:appendixSgrcolumnrownormalised}, i.e. the stars removed from the left panel of Fig. \ref{fig:Sgrcuts} using Eq. \ref{eq:Sgr_cut}.
The top row is just a reflection of the extent of the splines in the top row of Fig. \ref{fig:Sgrcuts} (with background or foreground stars removed via the $|\tilde{B}| < 15^\circ$ stream star requirement), but the bottom row shows how the selection clearly traces overdensities also in velocity space.
This is an important check that our stream star selection works well, as we apply it to our BHB sample to remove features in the halo that would, if left in, bias the velocity signature that we are trying to measure.

\section{The validity of our velocity field model}\label{sec:appendixmodelfittodata}

\begin{figure}
 \includegraphics[width=\columnwidth]{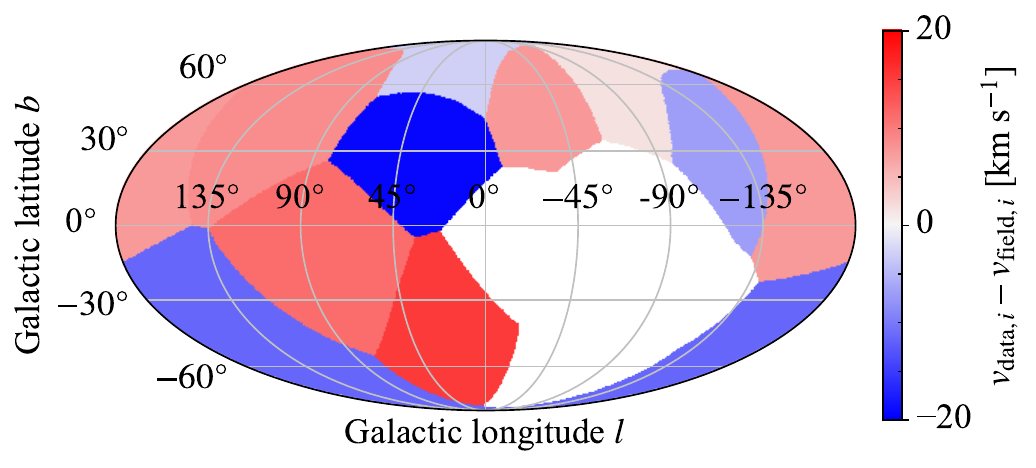}
 \caption{
 The residual of the observed BHB velocities and the velocities predicted for each star by the model given in \eqref{eq:vglobal}, using the parameters presented in Table \ref{tab:apexfittingresult}.
 } 
 \label{fig:appendixdatamodelresiduals}
 \end{figure}

We want to test how well the model $v_\text{field}$  we fit in Eq. \eqref{eq:vglobal} describes our data.
One thing to first check is how the residuals of $v_\text{field}$ and the observed velocities for each star are distributed on the sky.
As we can see in Fig. \ref{fig:appendixdatamodelresiduals} there is no pattern on the projected sky with the residuals between the model and the data.
This is an important check that we are not biasing ourselves by over- or underfitting a specific region on the sky.
We remind the reader that the mean error in each bin ranges from 7 to 18 \kms. 
This means that each bin has $v_\text{GSR}/\sigma < 2$, except for three bins that each have $v_\text{GSR}/\sigma = 2.7, 2.8$ and 3.1. 
All of these three bins are the ones in the Southern hemisphere.

\begin{figure}
 \includegraphics{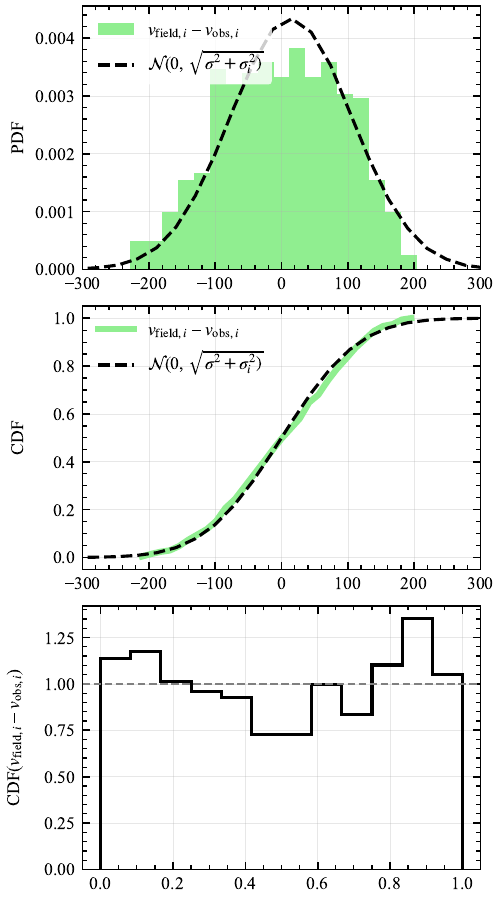}
 \caption{
 Comparison of the residuals of $v_{\text{field, } i} - v_{\text{obs, } i}$ for each individual star $i$ in our BHB sample, to the underlying, theoretical residual distribution $\mathcal{N}(0, \sqrt{ \sigma^2 + \sigma_i^2 })$ in the case of the model perfectly fitting the data. 
 \textit{Top row:} The probability density functions of the two samples. 
 \textit{Middle row:} The cumulative distribution functions of the two samples. 
 \textit{Bottom row:} The cumulative distribution function of $\mathcal{N}(0, \sqrt{ \sigma^2 + \sigma_i^2 })$, evaluated at $v_{\text{field, } i} - v_{\text{obs, } i}$.} 
 \label{fig:appendixmodeltests}
 \end{figure}

We want to further test the validity of our model given in Eq. \eqref{eq:vglobal}.
This is first done by doing a two-sample Kolmogorov-Smirnov test. 
The first sample is the residual of the predicted velocities of our stars given the model $v_{\text{field, } i}$\footnote{As the model in Eq. \eqref{eq:vglobal} takes values for $(l, b)$ as arguments, which we have for all our BHB stars, and we know the values of the parameters $v_\text{compr}$ and $v_\text{dipole}$ from the numerical fit, the model can predict the velocities of all stars.} and the observed velocities, $v_{\text{obs, } i}$, for every star $i$ in the sample.
These residuals should be normally distributed around a mean of 0, and have a standard deviation given by the standard deviation $\sigma$ on $v_\text{field}$ in Eq. \eqref{eq:vglobal} that the modelling outputs as the fifth parameter alongside the parameters in the equation, and the observational uncertainty on the velocity for star $i$, $\sigma_i$. 
We then construct the second sample by drawing 1,000 synthetic residuals from the distribution $\mathcal{N}(0, \sqrt{ \sigma^2 + \sigma_i^2 })$ for each star $i$, which means that because we have 675 stars in the sample, there will be 675,000 simulated residuals to compare to.
The distribution of the first sample is shown in green and of the second sample in black in the top row of Fig. \ref{fig:appendixmodeltests}.
The resulting p-value of the two-sample Kolmogorov-Smirnov test on these two samples is 0.12, so we conclude that the observed residuals are consistent with the $\mathcal{N}(0, \sqrt{ \sigma^2 + \sigma_i^2 })$ distribution.

If we again assume that $\mathcal{N}(0, \sqrt{ \sigma^2 + \sigma_i^2 })$ is the true distribution of our residuals, we can compute the cumulative distribution function (CDF) of it, given our observed residuals.
The CDF of the first sample and of the second sample is again seen in green and black respectively in the middle panel of Fig. \ref{fig:appendixmodeltests}.
A match between the two distributions should give us a uniform distribution, which is seen in the bottom panel of Fig. \ref{fig:appendixmodeltests}.
For this test, we see a distribution that somewhat deviates from a uniform distribution: there is a slight excess towards the tails of the distribution, and a dearth towards the middle.
This means that we have slightly fewer stars with a velocity residual around 0 \kms\ than expected from the Gaussian distribution, and instead have slightly more residuals that are high enough to lie towards the tails of the distribution.
This slight deviation, together with the p-value derived above, tells us that our model is a reasonable fit to the data, but could potentially be improved. 
We already know that the model used in this work is one-dimensional, only taking radial motion of the halo into account. 
It also does not vary with distance, but projects the velocities on to the celestial sphere.
Allowing for more (physically motivated) parameters of the model or variation with distance could improve it.

\section{Comparing the DESI and SDSS BHB samples}\label{sec:appendixDESISDSScomparison}

To compare the BHB selection performed by us with the SDSS BHB selection by \citet{Xue08_SDSSBHBSample}, which the sample in \citetalias{Xue11_SDSSDR8BHBs} is based on, we crossmatch the true, i.e. spectroscopically chosen, BHBs of one sample to the BHB candidates, i.e. before spectroscopic cleaning, to the other sample.
If the two selections created similar BHB catalogues, almost all of the candidates from one sample would be chosen as true BHBs by the other sample.
If this is not the case, this test allows us to test the purity and completeness of the two samples.
99 \pc\ of all true SDSS BHBs are also selected as BHBs by the DESI spectroscopic criteria given in Eq. \eqref{eq:Kielcuts}.
This means that their selection has a high purity, using the DESI definition of a BHB.
However, only 75 \% of all true DESI BHBs are selected as BHBs by the SDSS spectroscopic criteria.
This implies that either the DESI BHB selection is not pure, or that the SDSS BHB selection criteria are too conservative.

\begin{figure*}
 \includegraphics{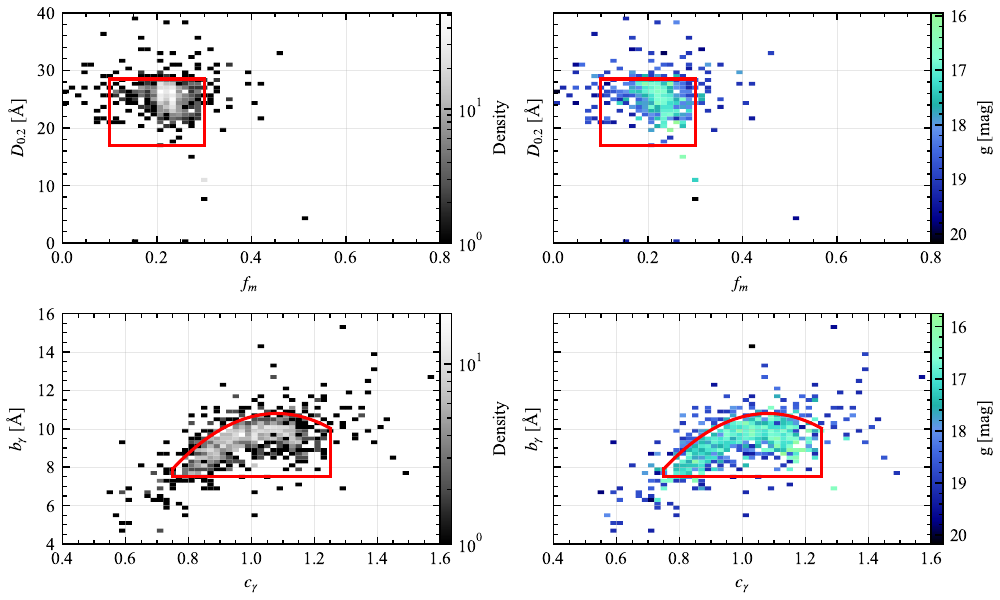}
 \caption{
 The DESI true BHB stars crossmatched with the \citet{Xue08_SDSSBHBSample} BHB candidates (i.e. stars that have passed the initial BHB candidate colour cut but not the spectroscopic BHB criteria), in the Balmer line selection spaces ($f_m$, $D_{0.2}$) (top row) and ($c_\gamma$, $b_\gamma$) (bottom row). 
 The regions in which stars are selected as true BHBs are shown in red, following the definitions in \citet{Xue08_SDSSBHBSample}. 
 The left-hand column is colour-coded by density and shows that a majority of DESI stars trace the BHB sequences defined by the red regions. 
 The right-hand column is colour-coded by SDSS $g$-magnitude and shows that most stars that are outside of the selection regions are faint. 
 The DESI stars do not trace any BS overdensities.
 } 
\label{fig:appendixX11DESIBHBcomparison}
\end{figure*}

We test this by checking where the true DESI BHBs lie in the Balmer line selection spaces ($f_m$, $D_{0.2}$) and ($c_\gamma$, $b_\gamma$) defined by \citet{Xue08_SDSSBHBSample} (see their Figs. 4 and 5).
This is shown in Fig. \ref{fig:appendixX11DESIBHBcomparison} (top and bottom row respectively), with the regions in which BHB stars are identified are marked in red.
In the left column of this figure, we clearly see that the majority of DESI BHBs trace the BHB sequences defined in the red regions, with random scatter outside of the selection regions. 
We test that this scatter, which in ($c_\gamma$, $b_\gamma$) appears to be quite large for $c_\gamma > 1.25$, is not due to contamination by rare blue stars, e.g. young main sequence stars, by plotting $c_\gamma$ as a function of both log($g$) and $T_\text{eff}$.
These are the parameters that the DESI selection in Eq. \eqref{eq:Kielcuts} is based on.
In the right-hand column, we see the same stars colour-coded by their SDSS $g$-magnitudes.
This shows that the majority of stars outside the selection boxes are faint, meaning that their spectra will have lower signal-to-noise ratios, and their stellar parameters are more difficult to determine.
This explains the random scatter around the selection boxes.
Since the BS sequences are not traced by the DESI true BHBs in the figure, we can conclude that the DESI BHB selection is indeed pure.
The reason that 25 \pc\ of DESI true BHBs are outside the SDSS selection regions is because the SDSS BHB cut is conservative.
This in turn implies that while the SDSS BHB sample is pure (since 99 \pc\ of them are also selected as BHBs by DESI), it is somewhat incomplete at faint magnitudes.

\section{Compare Sgr selections}\label{sec:appendixcompareSgrcuts}

A possible reason that we measure a different apex direction than \citetalias{Petersen21_LMCreflexmotionNature}, \citetalias{yaaqib24_Rashid} could be because we select Sgr stream stars differently.
Our selection is given in Eq. \eqref{eq:Sgr_cut}.
Their selection, explained in detail in \citetalias{Petersen21_LMCreflexmotionNature}, relies on Sgr latitude coordinate \Sgrlat\ and three-dimensional angular momentum components: stars should have |\Sgrlat| $< 20^\circ$ and angular momentum components $(L_x, L_y, L_z)$ within a 3,000 kpc \kms\ radius of the Sgr angular momentum components $(L_{x, \text{Sgr}}, L_{y, \text{Sgr}}, L_{z, \text{Sgr}}) = (605, -4515, -1267)$ kpc \kms.
We will refer to this selection as the $L$ Sgr selection.
Because it uses angular momentum components, there is a risk that this selection biases velocities, which is the main parameter to constrain when understanding the effect of the LMC on the halo.

\begin{figure}
 \includegraphics{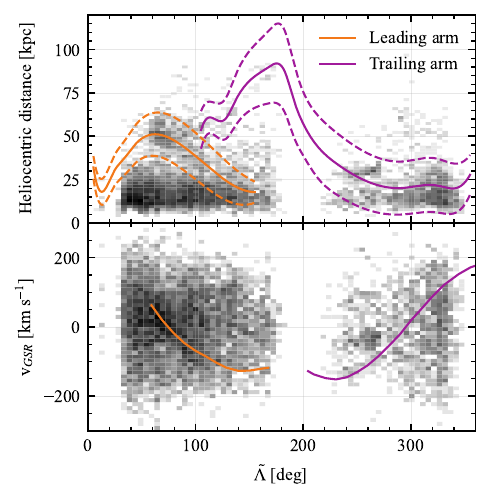}
 \caption{The DESI sample after removing Sgr stars based on the $L$ Sgr selection. Note the overdensity at the Sgr leading arm apocentre and Cetus-Palca, that is still in the sample.} 
 \label{fig:appendixSgrcomparisonLambda}
 \end{figure}

But more importantly, we see that when we apply their selection to our BHB sample, the removal of Sgr stars is incomplete.
This is seen in Fig. \ref{fig:appendixSgrcomparisonLambda}, which shows our BHB sample after removing stars based on the $L$ Sgr selection.
We can clearly see Sgr at the apocentre of the leading arm in the top panel, and Cetus-Palca almost perpendicular to the trailing arm.
This is reflected as overdensities in velocities.
Cetus-Palca stars have low enough Galactocentric distances to not enter into our analysis, however the Sgr leading arm apocentre stars will (remember that our analysis is based on Galactocentric distances $R>50$ kpc).
As we can see in the bottom panel of Fig. \ref{fig:appendixSgrcomparisonLambda}, the Sgr leading arm stars that are left in the sample have mainly $v_\text{GSR} > 0$ \kms.

\begin{figure}
 \includegraphics[width=\columnwidth]{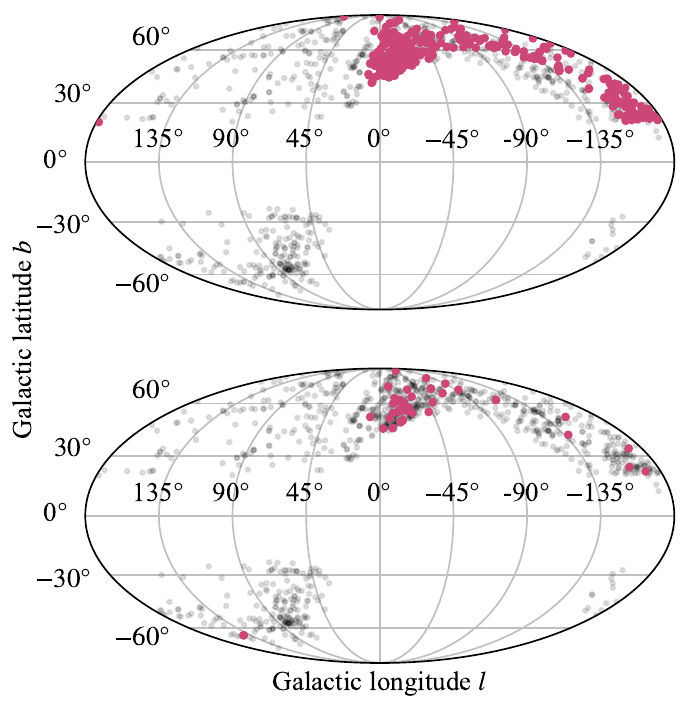}
 \caption{The on-sky distribution of the stars selected as Sgr members in purple by the method in Eq. \eqref{eq:Sgr_cut} (top panel) and the $L$ Sgr selection (bottom row), and the stars not selected in the background in grey.} 
 \label{fig:appendixSgrcomparisoncelestial}
 \end{figure}

In Fig. \ref{fig:appendixSgrcomparisoncelestial}, we see where the stars that are selected using this work's method in Eq. \eqref{eq:Sgr_cut} and the $L$ Sgr selection, and have $R>50$ kpc, are distributed in Galactic coordinates in purple (top and bottom panels, respectively).
Behind these stars, in grey, is the rest of the BHB sample (so that the grey stars in the bottom panel are the same stars seen in Fig. \ref{fig:appendixSgrcomparisonLambda}).
All of the Sgr-selected stars are in the Northern hemisphere (except for one using the $L$ selection method).
Because the $L$ selection is incomplete and leave Sgr stars in the sample, behind the Sgr stars in the bottom panel, we can still see an overdensity.
Most of these stars have positive velocities. 
Fitting the apex direction with these stream stars in the sample will rotate it towards a point further along the past orbit of the LMC, because the anti-apex direction in the Northern hemisphere, where the positive contributions from the dipole velocity peaks, will come closer to where these remnant Sgr stars are in the sample.
This leads to an apex direction that is closer to those reported in \citetalias{Petersen21_LMCreflexmotionNature}, \citetalias{yaaqib24_Rashid}.
If we take their data and apply their Sgr removal method to it, we recover their apex value.
If we instead apply our Sgr removal method in Eq. \eqref{eq:Sgr_cut} to their data, we recover an apex direction that is closer to ours.

For this reason, we believe that a non-negligible part of our tension with the results presented by \citetalias{Petersen21_LMCreflexmotionNature, yaaqib24_Rashid} can be explained by the difference in Sgr selection. 
However, we note that the apex direction derived using a MW satellite sample in \citetalias{Petersen21_LMCreflexmotionNature} is statistically consistent with that derived using their stellar sample, where the satellite sample should not be affected by Sgr stream debris, and so the Sgr selection cannot be the sole reason for our discrepant results.
That this difference in selection leads to shifts of the apex direction along the LMC past orbit is serendipitous because Sgr is on a polar orbit, similar to the future orbit of the LMC.

\section{Validating absence of halo rotation bias}\label{sec:appendixrotationbias}
The parametrization of the halo velocity field as given by Eq. \eqref{eq:vglobal} does not take the stellar halo's rotation into account.
A prograde halo rotation has been measured by \citet{Deason17_halospin} as ranging from 5--25 \kms\ in a Galactocentric radial range from 0--50 kpc, by \citet{Dillamore23_halorotation} as $\sim15$ \kms\ in the range 6--16 kpc, by \citetalias{yaaqib24_Rashid} as increasing from 11 \kms\ in a 20--30 kpc range to 24 \kms\ beyond 50 kpc, and by \citetalias{Petersen21_LMCreflexmotionNature} as being 20 \kms\ beyond 40 kpc.
\citet{Chandra24_LMCMW} measure a halo rotation that changes from retrograde to prograde as distance increases.
It was shown by \citetalias{Petersen21_LMCreflexmotionNature} that a rotating halo will create a similar pattern projected on the celestial sphere using line-of-sight velocities as a pure dipole signal (see their Extended Data Fig. 8), and that using a limited observational footprint may make it difficult to disentangle the rotation signal from a dipole signal.
They point out that the degeneracy between rotation and dipole signatures in velocity can be broken if transverse velocities are included in the fit, alongside radial velocities.
Considering that the DESI footprint does not cover the entire sky (see Fig. \ref{fig:onskydistribution}) and that we only utilise radial velocities in this work, we need to verify that the presence of rotation in the MW halo does not introduce a bias in our apex direction and that the halo rotation is not absorbed into $v_\mathrm{compr}$ or $v_\mathrm{dipole}$.

We test the potential bias in our parameters as given by Eq. \eqref{eq:vglobal} by comparing the fit using the L2M11 model (which does not contain ab initio rotation) seen in Sec. \ref{sec:modelcomparison}, with a new fit after rotation has been projected onto the radial velocities along the line-of-sight.
This is done first for all stellar particles past 50 kpc, and second we compare the fits in different distance bins, since we expect that the effect of rotation should increase as Galactocentric distances decrease.
For the first test, we apply three different rotations: --20 \kms\ from \citetalias{Petersen21_LMCreflexmotionNature}, --24 \kms\ from \citetalias{yaaqib24_Rashid}, and 12 \kms\ at 40 kpc to --10 \kms\ at 120 kpc with a linear variation in between, from \citet{Chandra24_LMCMW}.
The resulting apex directions are shown in Table \ref{tab:rotationbias}, together with the angular separation between the case of no rotation (as seen in Sec. \ref{sec:modelcomparison}) and the different cases of added rotation, which we denote $\theta$. A larger value of $\theta$ would mean that the assumed rotation would have a larger effect on the derived apex direction, and the opposite is true for a smaller $\theta$.
However, from the range of $\theta$ values we see that the projection of rotation onto the line-of-sight shifts the apex direction very little, and the angular differences are much smaller than the errors on $(l_\text{apex}, b_\text{apex})$ from the DESI BHB sample, given in Table \ref{tab:apexfittingresult}.
For this reason, we believe that the rotation of the halo does not affect our velocity field measurement.

\begin{table}
	\centering
	\caption{
        The resulting $(l_\text{apex}, b_\text{apex})$ from redoing the fits of Eq. \eqref{eq:vglobal} on the L2M11 by \citet{Vasiliev24_LMCMWmodels}, with varying amount of halo rotation projected onto the radial velocity, and the angular separation $\theta$ between the unmodified apex direction and those with added rotation.}
        \label{tab:rotationbias}
	\begin{tabular}{lcc}
		\hline
		Rotation source & $(l_\text{apex}, b_\text{apex})$ [deg] &$\theta$ [deg] \\
		\hline
		No rotation                                   & $(-25.3, -78.9)$ & -- \\ 
		\citetalias{Petersen21_LMCreflexmotionNature} & $(-4.3, -79.9)$   & 4.0 \\
		\citetalias{yaaqib24_Rashid}                  & $(0.2, -79.9)$  & 4.8 \\
		\citet{Chandra24_LMCMW}                       & $(-27.1, -78.8)$ & 0.4 \\
		\hline
	\end{tabular}
\end{table}

\begin{figure}
 \includegraphics{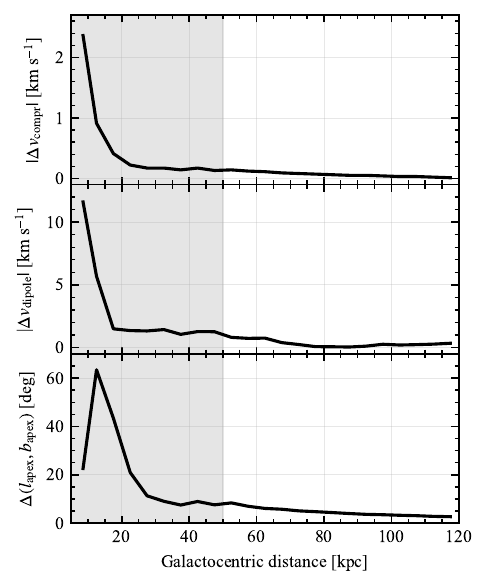}
 \caption{
 The absolute value of the difference in the compression velocity $v_\text{compr}$ (top row), dipole velocity $v_\text{dipole}$ (middle row) and the apex direction $(l_\text{apex}, b_\text{apex})$ (bottom row) for the L2M11 model without rotation and rotation from \citet{Dillamore23_halorotation} in the Galactocentric radial range 6--20 kpc and from \citetalias{yaaqib24_Rashid} beyond 20 kpc.
 The distance range used, $6 < R < 120$ kpc, is chosen to match that of the BHB distribution, see Fig. \ref{fig:distancedistributions}.
 } 
 \label{fig:appendixrotationbias}
 \end{figure}

For the second test, we bin the L2M11 model in 5 kpc distance bins that range from 6 kpc to 120 kpc, to match the distance distribution of our BHB sample, and repeat the fit of Eq. \eqref{eq:vglobal} in each distance bin, for both the unmodified case and when the rotation from \citetalias{yaaqib24_Rashid} has been added to the line-of-sight.
We choose their rotation measurement as their rotation amplitude in the outer halo is the highest, and it thus represents a worst-case scenario. 
This test should inform us of the magnitude of the bias introduced by rotation.
Since their rotation is only measured down to a Galactocentric distance of 20 kpc, we supplement this with the rotation from \citet{Dillamore23_halorotation} in the range 6 to 20 kpc.
The differences between the unmodified case and the rotation-added case can be seen in Fig. \ref{fig:appendixrotationbias}.
These plots show that the effects of adding rotation indeed are the largest for smaller distance bins, and at $R < 30$ kpc, the effects are substantial.
This means that in the inner halo, we cannot accurately measure the dipole, as the results from fitting Eq. \eqref{eq:vglobal} will return biased results.
However, beyond 50 kpc, where we perform our apex direction fits, the effects of rotation are completely negligible, and well within the measurement uncertainties of the fits in Table \ref{tab:apexfittingresult}: the $v_\mathrm{compr}$ amplitude changes are less than 0.4 \kms, the $v_\mathrm{dipole}$ amplitude changes are less than 2 \kms, and the angular differences less than 10 degrees.
At $R = 59.9$ kpc, which is the median distance of our sample when $R>50$ kpc, the difference in $(l_\text{apex}, b_\text{apex})$ is 6.6 degrees.

These tests show that since the bias from halo rotation being projected onto the line-of-sight and thus our radial velocities is significantly smaller than the measurement uncertainties of our fit of the halo velocity field, we assume that a rotating halo does not bias the results presented in this paper.
The rotation bias could be even smaller than presented in the second test, which assumed the highest amplitude rotation in literature \citepalias{yaaqib24_Rashid}.
It is also clear that the effects of rotation are significant in the inner parts of the MW, which confirms the importance of using distant halo tracers to understand the effects of the MW's reflex motion on the halo velocity field.

\vspace{1em}

\noindent $^{1}$ \textit{Institute for Astronomy, University of Edinburgh, Royal Observatory, Blackford Hill, Edinburgh EH9 3HJ, UK}\\
$^{2}$ \textit{Institute of Astronomy, University of Cambridge, Madingley Road, Cambridge CB3 0HA, UK}\\
$^{3}$ \textit{Institute for Computational Cosmology, Department of Physics, Durham University, South Road, Durham DH1 3LE, UK}\\
$^{4}$ \textit{Department of Astronomy \& Astrophysics, University of Toronto, Toronto, ON M5S 3H4, Canada}\\
$^{5}$ \textit{Department of Astronomy, University of Michigan, Ann Arbor, MI 48109, USA}\\
$^{6}$ \textit{Steward Observatory, University of Arizona, 933 N, Cherry Ave, Tucson, AZ 85721, USA}\\
$^{7}$ \textit{Department of Physics and Astronomy, Carleton College, 1 North College Street, Northfield, MN 55057, United States}\\
$^{8}$ \textit{Center for Astrophysics $|$ Harvard \& Smithsonian, 60 Garden Street, Cambridge, MA 02138, USA}\\
$^{9}$ \textit{NSF NOIRLab, 950 N. Cherry Ave., Tucson, AZ 85719, USA}\\
$^{10}$ \textit{Instituto de Astrof\'{\i}sica de Canarias, C/ V\'{\i}a L\'{a}ctea, s/n, E-38205 La Laguna, Tenerife, Spain}\\
$^{11}$ \textit{Lawrence Berkeley National Laboratory, 1 Cyclotron Road, Berkeley, CA 94720, USA}\\
$^{12}$ \textit{Physics Dept., Boston University, 590 Commonwealth Avenue, Boston, MA 02215, USA}\\
$^{13}$ \textit{Departamento de Astrof\'{\i}sica, Universidad de La Laguna (ULL), E-38206, La Laguna, Tenerife, Spain}\\
$^{14}$ \textit{Department of Physics \& Astronomy, University College London, Gower Street, London, WC1E 6BT, UK}\\
$^{15}$ \textit{Department of Physics and Astronomy, The University of Utah, 115 South 1400 East, Salt Lake City, UT 84112, USA}\\
$^{16}$ \textit{Instituto de F\'{\i}sica, Universidad Nacional Aut\'{o}noma de M\'{e}xico,  Cd. de M\'{e}xico  C.P. 04510,  M\'{e}xico}\\
$^{17}$ \textit{Institut de F\'{i}sica d’Altes Energies (IFAE), The Barcelona Institute of Science and Technology, Campus UAB, 08193 Bellaterra Barcelona, Spain}\\
$^{18}$ \textit{Departamento de F\'isica, Universidad de los Andes, Cra. 1 No. 18A-10, Edificio Ip, CP 111711, Bogot\'a, Colombia}\\
$^{19}$ \textit{Observatorio Astron\'omico, Universidad de los Andes, Cra. 1 No. 18A-10, Edificio H, CP 111711 Bogot\'a, Colombia}\\
$^{20}$ \textit{Institut d'Estudis Espacials de Catalunya (IEEC), 08034 Barcelona, Spain}\\
$^{21}$ \textit{Institute of Cosmology and Gravitation, University of Portsmouth, Dennis Sciama Building, Portsmouth, PO1 3FX, UK}\\
$^{22}$ \textit{Institute of Space Sciences, ICE-CSIC, Campus UAB, Carrer de Can Magrans s/n, 08913 Bellaterra, Barcelona, Spain}\\
$^{23}$ \textit{Sorbonne Universit\'{e}, CNRS/IN2P3, Laboratoire de Physique Nucl\'{e}aire et de Hautes Energies (LPNHE), FR-75005 Paris, France}\\
$^{24}$ \textit{Instituci\'{o} Catalana de Recerca i Estudis Avan\c{c}ats, Passeig de Llu\'{\i}s Companys, 23, 08010 Barcelona, Spain}\\
$^{25}$ \textit{Department of Physics and Astronomy, Siena College, 515 Loudon Road, Loudonville, NY 12211, USA}\\
$^{26}$ \textit{Instituto de Astrof\'{i}sica de Andaluc\'{i}a (CSIC), Glorieta de la Astronom\'{i}a, s/n, E-18008 Granada, Spain}\\
$^{27}$ \textit{Departament de F\'isica, EEBE, Universitat Polit\`ecnica de Catalunya, c/Eduard Maristany 10, 08930 Barcelona, Spain}\\
$^{28}$ \textit{Department of Physics and Astronomy, Sejong University, Seoul, 143-747, Korea}\\
$^{29}$ \textit{CIEMAT, Avenida Complutense 40, E-28040 Madrid, Spain}\\
$^{30}$ \textit{Department of Physics, University of Michigan, Ann Arbor, MI 48109, USA}\\
$^{31}$ \textit{National Astronomical Observatories, Chinese Academy of Sciences, A20 Datun Rd., Chaoyang District, Beijing, 100012, P.R. China}\\


\bsp	
\label{lastpage}
\end{document}